\documentclass[sigplan,screen]{acmart}

\AtBeginDocument{%
  }

\usepackage{balance}
\usepackage{fixltx2e}
\usepackage{enumitem}
\usepackage{booktabs}
\usepackage{multirow}
\usepackage{multicol,makecell}
\usepackage{xspace}
\usepackage{graphicx}
\usepackage{xcolor}
\usepackage{fancyhdr}
\usepackage[normalem]{ulem}
\usepackage[bottom]{footmisc}
\usepackage{courier}
\usepackage{dblfloatfix}
\usepackage{longfbox}
\usepackage{caption}
\usepackage[]{hyperref}
\usepackage[colorinlistoftodos,prependcaption,textsize=scriptsize]{todonotes} 

\definecolor{pred}{rgb}{0.7843, 0.0039, 0.3137} 
\definecolor{darknavy}{rgb}{0, 0, 0.5}

\newcommand\rev[1]{{\color{black}{#1}}}

\newcommand{\fig}[1]{{Figure~#1}\xspace}
\newcommand{\figs}[1]{{Figures~#1}\xspace}
\newcommand{\sect}[1]{{\S#1}\xspace}

\newcommand{\three}{$\tau_\text{99.9P}$\xspace}
\newcommand{\four}{$\tau_\text{99.99P}$\xspace}
\newcommand{\six}{$\tau_\text{99.9999P}$\xspace}

\newcommand\proposal{\textsc{Aero}\xspace}
\newcommand\impl{\textsc{AeroFTL}\xspace}
\newcommand\felpfull{\emph{\underline{F}ail-bit-count-based \underline{E}rase \underline{L}atency \underline{P}rediction (FELP)}\xspace}
\newcommand\felp{\text{FELP}\xspace}
\newcommand\ispe{\text{ISPE}\xspace}

\newcommand\inum[1]{(\textit{#1})\xspace}
\newcommand{\head}[1]{{\noindent\textbf{#1.}\xspace}}
\newcommand{\vth}{V$_{\text{TH}}$\xspace}
\newcommand{\vref}{V$_{\text{REF}}$\xspace}

\newcommand{\vverify}{V$_{\text{VERIFY}}$\xspace}
\newcommand{\verase}{V$_\text{ERASE}$\xspace}
\newcommand{\verasei}[1]{V$_{\text{ERASE}}$($#1$)\xspace}
\newcommand{\dverase}{$\Delta$V$_\text{ISPE}$\xspace}

\newcommand{\usec}{\textmu{}s\xspace} 

\newcommand{\tr}{\texttt{tR}\xspace}
\newcommand{\tprog}{\texttt{tPROG}\xspace}
\newcommand{\tbers}{\texttt{tBERS}\xspace}
\newcommand{\mtbers}{$m_\texttt{tBERS}$\xspace}
\newcommand{\mtepi}[1]{$m_\texttt{tEP}$($#1$)\xspace}
\newcommand{\mtep}{\mtepi{N_\text{ISPE}}}

\newcommand{\nispe}{$N_\text{ISPE}$\xspace}

\newcommand{\tep}{\texttt{tEP}\xspace} 
\newcommand{\tvr}{\texttt{tVR}\xspace} 
\newcommand{\tshallow}{\texttt{tSE}\xspace} 
\newcommand{\vr}{\text{VR}\xspace}
\newcommand{\ep}{\text{EP}\xspace}
\newcommand{\epi}[1]{EP($#1$)\xspace} 
\newcommand{\vri}[1]{VR($#1$)\xspace} 

\newcommand{\dpes}{\text{DPES}\xspace} 
\newcommand{\iispe}{$i$-ISPE\xspace} 

\newcommand{\mrber}{$M_\text{RBER}$\xspace}
\newcommand{\mrberx}[1]{$M_\text{RBER}$($#1$)\xspace}
\newcommand{\nfail}[1]{$F$($#1$)\xspace} 
\newcommand{\fpass}{$F_\text{PASS}$\xspace}
\newcommand{\fhigh}{$F_\text{HIGH}$\xspace}
\newcommand{\nloop}{$N_\text{ISPE}$\xspace}

\newcommand{\basessd}{\textsf{Baseline}\xspace}
\newcommand{\aerossd}{\textsf{AERO$_\textsf{CONS}$}\xspace}
\newcommand{\dpesssd}{\textsf{DPES}\xspace}
\newcommand{\aerooptssd}{\textsf{AERO}\xspace}
\newcommand{\iispessd}{\textsf{i-ISPE}\xspace}

\newcommand{\degreec}[1]{#1$^\circ$C\xspace}

\DeclareRobustCommand\bcirc[1]{\tikz[baseline=(char.base)]{
           \node[shape=circle,draw,inner sep=0pt,fill=black, text=white] (char) {#1};}}
\DeclareRobustCommand\wcirc[1]{\tikz[baseline=(char.base)]{
           \node[shape=circle,draw,inner sep=0pt,fill=white] (char) {#1};}}

\copyrightyear{2024}
\acmYear{2024}
\setcopyright{acmlicensed}
\acmConference[ASPLOS '24]{29th ACM International Conference on Architectural Support for Programming Languages and Operating Systems, Volume 1}{April 27-May 1, 2024}{La Jolla, CA, USA}
\acmBooktitle{29th ACM International Conference on Architectural Support for Programming Languages and Operating Systems, Volume 1 (ASPLOS '24), April 27-May 1, 2024, La Jolla, CA, USA}
\acmPrice{15.00}
\acmDOI{10.1145/3620666.3651341}
\acmISBN{979-8-4007-0386-7/24/04}
\title[\proposal: Adaptive Erase Operation for Improving Performance and Lifetime of Modern SSDs]{\proposal: Adaptive Erase Operation for Improving Lifetime and Performance of Modern NAND Flash-Based SSDs}

\author{Sungjun Cho}
\email{allencho1222@postech.ac.kr}
\orcid{0000-0002-8609-6183}
\affiliation{%
  \institution{POSTECH}
  \country{Republic of Korea}
}

\author{Beomjun Kim}
\email{beomjun0816@knu.ac.kr}
\orcid{0000-0003-1211-5641}
\affiliation{%
  \institution{Kyungpook National University}
  \country{Republic of Korea}
}

\author{Hyunuk Cho}
\email{gusdnr9779@postech.ac.kr}
\orcid{0000-0002-1554-1884}
\affiliation{%
  \institution{POSTECH}
  \country{Republic of Korea}
}

\author{Gyeongseob Seo}
\email{syhbong9@knu.ac.kr}
\orcid{0009-0008-9901-9874}
\affiliation{%
  \institution{Kyungpook National University}
  \country{Republic of Korea}
}

\author{Onur Mutlu}
\email{omutlu@gmail.com}
\orcid{0000-0002-0075-2312}
\affiliation{%
  \institution{ETH Z\"urich}
  \country{Switzerland}
}
  
\author{Myungsuk Kim}
\email{ms.kim@knu.ac.kr}
\orcid{0000-0002-8667-3198}
\affiliation{%
  \institution{Kyungpook National University}
  \country{Republic of Korea}
}

\author{Jisung Park}
\email{jisung.park@postech.ac.kr}
\orcid{0000-0002-1826-9003}
\affiliation{%
  \institution{POSTECH}
  \country{Republic of Korea}
}
\date{March 2024}
\renewcommand{\shortauthors}{Cho et al.}
\begin{document}
\captionsetup{font=small}

\begin{abstract}
This work investigates a new erase scheme in NAND flash memory to improve the lifetime and performance of modern solid-state drives (SSDs).
In NAND flash memory, an erase operation applies a high voltage (e.g., $>$~20~V) to flash cells for a long time (e.g., $>$~3.5~ms), which degrades cell endurance and potentially delays user I/O requests.
While a large body of prior work has proposed various techniques to mitigate the negative impact of erase operations, no work has yet investigated how erase latency should be set to fully exploit the potential of NAND flash memory; most existing techniques use a fixed latency for every erase operation which is set to cover the worst-case operating conditions.
To address this, we propose \proposal (\underline{A}daptive \underline{ER}ase \underline{O}peration), a new erase scheme that dynamically adjusts erase latency to be just long enough for reliably erasing target cells, depending on the cells' current erase characteristics.
\proposal accurately predicts such near-optimal erase latency based on the number of fail bits during an erase operation.
To maximize its benefits, we further optimize \proposal in two aspects.
First, at the beginning of an erase operation, \proposal attempts to erase the cells for a short time (e.g., 1~ms), which enables \proposal to always obtain the number of fail bits necessary to accurately predict the near-optimal erase latency.
Second, \proposal~aggressively yet safely reduces erase latency by leveraging a large reliability margin present in modern SSDs.
We demonstrate the feasibility and reliability of \proposal using 160 real 3D NAND flash chips, showing that it enhances SSD lifetime over the conventional erase scheme by 43\% without change to existing NAND flash chips.
Our system-level evaluation using eleven real-world workloads shows that an \proposal-enabled SSD reduces read tail latency by 34\% on average over a state-of-the-art technique.
\end{abstract}


\maketitle 
\thispagestyle{empty}
\section{Introduction}
\label{sec:intro}
NAND flash memory is the prevalent memory technology in architecting modern storage systems.
NAND flash-based solid-state drives (SSDs) offer various advantages over traditional hard disk drives, such as high performance and small form factor, while providing high device capacity (e.g., several tens of terabytes per SSD~\cite{samsung, hynix, micron-ssd, wd}).
Although there exist several emerging non-volatile memory technologies (e.g., \cite{wong-ieee-2010, zangeneh-vlsi-2013, aggarwal-flashmemorysummit-2019, kawashima-2009}),
NAND flash memory is (expected to be) the predominant technology for storage systems to meet the increasing capacity demands of modern data-intensive applications.

The efficiency of erase operation performed by NAND flash chips significantly affects SSD lifetime and I/O performance due to two key reasons.
First, the high erase voltage physically damages flash cells.
After experiencing a certain number of program and erase (P/E) cycles, a flash cell cannot reliably store data, thereby limiting SSD lifetime.
Second, erase latency is significantly higher (e.g., 3.5~ms~\cite{cho-isscc-2021, kim-isscc-2022}) than read and write latencies (e.g., 40~\usec and 350~\usec, respectively~\cite{cho-isscc-2021}) due to the orders-of-magnitude larger granularity of erase operations (i.e., a block) compared to read and program operations (i.e., a page).
From the lifetime perspective, such a long latency leads an erase operation to have a much higher impact on cell endurance compared to a program operation that also applies a high voltage to target cells~\cite{hong-fast-2022}.
From the I/O performance perspective, an erase operation often delays I/O requests for a long time (e.g., several milliseconds), which significantly increases the SSD tail latency~\cite{wu-fast-2012, kim-fast-2019}.

An erase operation in modern SSDs often requires \emph{multiple erase loops}, which further increases the performance/lifetime impact of erase operations.
A flash cell becomes more difficult to erase as it experiences more P/E cycles~\cite{cai-ieee-2017, kim-pe-2021, hong-fast-2022}, so an erase operation with the default erase voltage may fail to sufficiently erase every cell in the block, which we call an \emph{erase failure}.
To ensure data reliability, modern NAND flash memory commonly employs the \ispe (\underline{I}ncremental \underline{S}tep \underline{P}ulse \underline{E}rasure) scheme~\cite{ispe}; when an erase failure occurs, the \ispe scheme \emph{retries} an erase loop with progressively higher voltages until it can successfully erase all the cells in the block.
We find that an erase failure occurs quite frequently in modern SSDs with density-optimized NAND flash chips, which aggravates the wear-out and long tail latency problems.
For example, our characterization study using 160 real 3D triple-level cell (TLC) NAND flash chips shows that every erase operation requires at least two erase loops (up to five loops) after the target block experiences 2K P/E cycles.

Even though a large body of prior work~\cite{lee-ispass-2011,cui-date-2018,kang-dac-2018,shahidi-sc-2016,choi-hpdc-2018,guo-ipdps-2017,kang-cm-2017,pan-hotstorage-2019,lee-tcad-2013, murugan-msst-2011,li-msst-2019,dh-tcad-2022,jeong-hotstorage-2013, jeong-fast-2014, jeong-tc-2017, wu-fast-2012,kim-fast-2019, ispe} has investigated various optimizations to mitigate the negative impact of erase operations, no work has yet investigated \emph{how erase latency should be set} to fully exploit the potential of NAND flash memory.
To be specific, most existing techniques use a \emph{fixed} latency for \emph{every} erase operation which is set by the manufacturers at design time based on the \emph{worst-case} operating conditions.
Like other memory technologies (e.g., DRAM), however, modern NAND flash memory also exhibits high process variation, which introduces significant differences in physical characteristics across flash cells~\cite{shim-micro-2019, luo-acm-2018, wang-acm-2017, chen-dac-2017, hung-ssc-2015, yen-hpca-2022, li-micro-2020, cai-date-2013, cai-dsn-2015}. 
For example, our real-device characterization study in \sect{\ref{sec:device_characterization_study}} shows that it is possible to completely and reliably erase a majority of blocks (e.g., 79\%-90\%) with much lower latency (e.g., by 17\%-29\%) than the default erase latency under many operating conditions.
This means that flash cells frequently suffer from more erase-induced damage than needed, which, in turn, degrades both SSD lifetime and I/O performance.

\textbf{Our goal} in this work is to improve the lifetime and performance of modern SSDs by mitigating the negative impact of erase operations.
To this end, we propose \emph{\proposal (\underline{A}daptive \underline{ER}ase \underline{O}peration)}, a new block erasure mechanism for NAND flash memory which is redesigned based on thorough characterization of real 3D TLC NAND flash chips. 
\textbf{The key idea} of \proposal is to dynamically adjust erase latency to be \emph{just long enough} for reliably erasing the target cells, depending on the cells' current erase characteristics.

Our key idea is simple, but it is challenging to accurately predict the minimum latency for a block in modern NAND flash memory.
Even though prior work~\cite{chen-trans-2018,lue-iedm-2015} has demonstrated a strong correlation between the erase latency and P/E-cycle count (PEC) of a block, PEC alone is insufficient for accurate prediction of the minimum erase latency due to high process variation across blocks.
Our real-device characterization results from 160 real 3D TLC NAND flash chips (\sect{\ref{sec:device_characterization_study}}) show high erase-latency variations even across blocks with the same PEC, e.g., a standard deviation of 2.7~ms in the erase latency across blocks with 3.5K PEC.

To address the challenge, \proposal introduces \felpfull that accurately predicts near-optimal latency for an erase loop based on the number of fail bits that occur in the previous loop.
At the end of each erase loop, the \ispe scheme senses all the cells in the target block simultaneously and counts the number of \emph{fail bits}, i.e.,~the number of bitlines that contain one or more insufficiently-erased cells, so as to perform another loop if the fail-bit count is larger than a threshold.
We find that the fail-bit count can be an accurate proxy for the minimum latency of the \emph{next} loop, as the more sufficiently the cells are erased, the lower the fail-bit count.
We construct a model between the fail-bit count in an erase loop and the minimum erase latency required for the next loop, which enables \proposal to safely reduce erase latency depending on the block's characteristics.  

We further optimize \proposal in two aspects.
First, we enable \proposal to also optimize single-loop erase operations by performing the first erase loop in two steps: \inum{i}~shallow erasure, which applies the erase voltage for a reduced amount of time (e.g., 1~ms), and \inum{ii}~remainder erasure, which completes erasing the block only with the necessary time determined based on the fail-bit count in the shallow erasure.
Second, we leverage the high error-correction capability of modern SSDs to further reduce erase latency without compromising reliability.
To cope with the low reliability of NAND flash memory, modern SSDs commonly adopt sophisticated error-correction codes (ECC),
which leads to a large ECC-capability margin in many cases~\cite{cai-ieee-2017, park-asplos-2021}.
Aggressive erase-latency reduction would inevitably cause insufficient erasure of some flash cells, potentially incurring more bit errors.
To ensure data reliability, \proposal carefully reduces erase latency for certain operating conditions we find via extensive real-device characterization.

\proposal provides high lifetime and performance benefits with small overheads.
\proposal requires only small changes to existing SSD firmware or controller but no modification to NAND flash chips, thereby achieving high applicability and practicality.
Our real-device characterization and system-level evaluation with a state-of-the-art SSD simulator~\cite{tavakkol-fast-2018} show that \proposal enhances SSD lifetime by 13\% and reduces the 99.9999th percentile read latency by 34\% on average compared to state-of-the-art techniques~\cite{ispe, jeong-hotstorage-2013, jeong-fast-2014, jeong-tc-2017}.

The key contributions of this work are as follows:
\begin{itemize}[leftmargin=*, noitemsep, topsep=0pt]
    \item To our knowledge, this work is the first to identify a new opportunity to safely reduce the erase latency without compromising reliability, which significantly mitigates the negative impact of erase operations.
    \item We introduce \proposal, a new block erasure mechanism that dynamically adjusts the erase latency based on varying erase characteristics of the target flash blocks.
    \item We validate the feasibility and reliability of \proposal via rigorous characterization of real 3D NAND flash chips.
    \item We evaluate the effectiveness of \proposal using real-world workloads, showing large lifetime and performance benefits over state-of-the-art techniques~\cite{ispe, jeong-hotstorage-2013, jeong-fast-2014, jeong-tc-2017}.
\end{itemize}
\section{Background}\label{sec:background}
We provide a brief background on NAND flash memory necessary to understand the rest of the paper.

\subsection{NAND Flash Basics}\label{ssec:background:basic}
\begin{figure}[h]
     \centering
     \includegraphics[width=\linewidth]{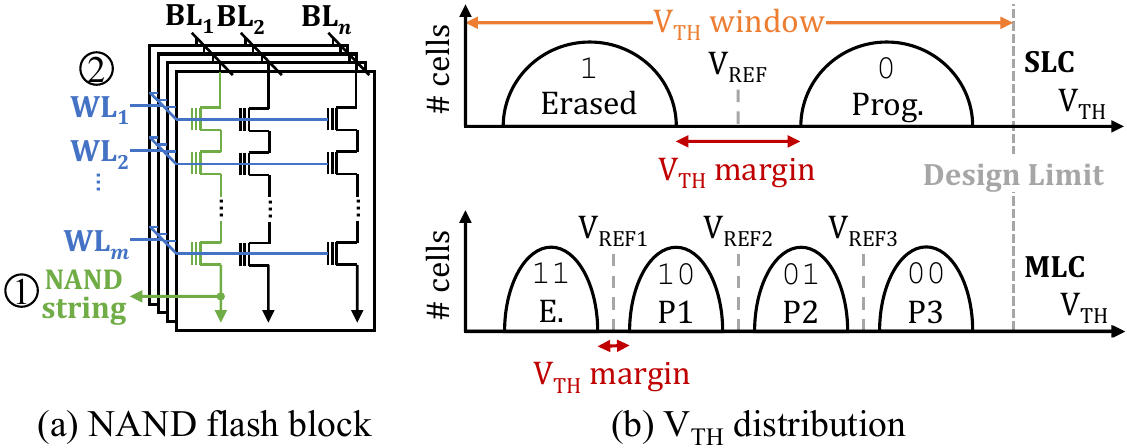}
     \caption{Overview of 3D NAND flash memory.}
     \label{fig:nand_mem}
\end{figure}

\head{NAND Flash Organization}
\fig{\ref{fig:nand_mem}a} depicts the hierarchical organization of 3D NAND flash memory.
A set of flash cells form a \emph{NAND string} (\wcirc{1} in \fig{\ref{fig:nand_mem}a}) that is connected to a bitline~(BL), and NAND strings of different BLs compose a \emph{block}.
The control gate of each cell at the same vertical locations in a block is connected to a wordline (WL) in parallel (\wcirc{2}),  so all the cells at the same WL concurrently operate.

\head{Data Storage Mechanism}
A flash cell stores bit data as a function of its threshold voltage (\vth) level that highly depends on the amount of charge in the cell's charge trap;
the more the electrons in the charge trap layer, the higher the cell's \vth level.
To increase the storage density, a flash cell can store multiple bits by controlling its \vth level more precisely, which is called \emph{multi-level cell (MLC)} technology.
\fig{\ref{fig:nand_mem}b} compares the \vth distribution in NAND flash memory when it stores one-bit data per cell using two \vth \emph{states} (top) and when it stores two bits per cell using four \vth{} states (bottom).

\head{NAND Flash Operations}
There are three basic operations to enable access to NAND flash memory: \inum{i}~program, \inum{ii}~read, and \inum{iii} erase operations.
A program operation \emph{increases} a target cell's \vth level by applying a high program voltage  
(e.g., $>$ 20 V).
A read operation determines a cell's \vth level (i.e., the stored data) by applying a read-reference voltage \vref to the cell's control gate; depending on its \vth level, the cell operates as either an on-switch (\vref $>$ \vth) or an off-switch (\vref $<$ \vth).
Since a number of flash cells (e.g.,~\mbox{$>2^{17}$}) share a WL, NAND flash memory performs read and program operations at a \emph{page} granularity (e.g., 16 KiB).

An erase operation \emph{decreases} a target cell's \vth level by applying a high erase voltage \verase{} (e.g., $>$ 20~V) to the substrate.
NAND flash memory performs an erase operation at \emph{block} granularity, which causes erase latency \tbers to be significantly higher than program latency \tprog and read latency \tr (e.g., 
$\texttt{tBERS}=3.5$~ms, $\texttt{tPROG}=400$~\usec, and $\texttt{tR}=40$~\usec~\cite{cho-isscc-2021}) but enables high erase bandwidth by concurrently erasing a large number of pages (e.g., more than 2K pages~\cite{cho-isscc-2021}) at once.
As a program operation can \emph{only} increase a cell's \vth level, a page needs to be erased first to program data, which is called the \emph{erase-before-write} property.

\subsection{NAND Flash Reliability}
\label{ssec:bg_nand_reliability}
NAND flash memory is highly error-prone due to its imperfect physical characteristics.
A flash cell leaks its charge (i.e.,~its \vth level decreases) over time, which is called retention loss. 
Reading or programming cells slightly increases the \vth levels of other cells in the same block (e.g., read/program disturbance~\cite{cai-dsn-2015, ha-tcad-2016, liu-asplos-2021, cai-iccd-2013, park-dac-2016, kim-dac-2017, cai-hpca-2017}).
If a cell's \vth level shifts beyond the \vref{} values (i.e., to adjacent \vth ranges corresponding to different bit values), reading the cell causes a bit error.

There are two major factors that significantly increase the raw bit-error rate (RBER) of NAND flash memory.
First, the high voltage used in program/erase operations physically damages flash cells, making the cells more error-prone. 
Second, the MLC technique increases RBER because packing more \vth states within a limited \vth window narrows the margin between adjacent \vth{} states as shown in \fig{\ref{fig:nand_mem}b}.

To ensure data reliability, it is common practice to employ strong error-correction codes (ECC). 
ECC stores redundant bits called ECC parity, which enables detecting and correcting raw bit errors in the codeword. 
To cope with the high RBER of modern NAND flash memory, modern SSDs use sophisticated ECC that can correct several tens of raw bit errors per 1-KiB data (e.g., low-density parity-check (LDPC) codes~\cite{ldpc}).
\section{Motivation}\label{sec:motivation}
We discuss \inum{i}~the importance of optimizing erase operations and \inum{ii}~limitations of existing techniques.
Table~\ref{tab:terminology} in Appendix summarizes new terminologies defined in this work.
\subsection{Negative Impact of Erase Operations}
\label{ssec:motiv_impact}
Erase operation significantly affects both SSD lifetime and I/O performance.
First, an erase operation is the major source that limits SSD lifetime.
The high voltage used in program and erase operations damages flash cells, which makes a block \emph{unusable} after experiencing a certain number of program and erase (P/E) cycles (e.g., 5K P/E cycles~\cite{kim-pe-2021}) because it can no longer meet the retention requirements for a non-volatile storage medium (e.g., 1-year retention at 30$^\circ$C~\cite{jedec-1y}).
Program operation also applies a high voltage to flash cells, but prior work has experimentally demonstrated that erase operation accounts for 80\% of cell stress due to the significantly longer erase latency compared to program latency~\cite{hong-fast-2022}.

Second, the long erase latency often increases the tail latency of user reads significantly, which is critical for data-center applications~\cite{wu-fast-2012, kim-fast-2019}. 
In fact, the impact of erase operations on the \emph{average} I/O performance is trivial in modern SSDs~\cite{kim-fast-2019}, since they occur much less frequently compared to read and program operations.
For example, a block in modern 3D NAND flash memory consists of more than 2K pages~\cite{kim-isscc-2022, yuh-isscc-2022, kim-isscc-2023}, so one erase operation occurs after at least 2K page writes (and even more page reads potentially).
However, an erase operation may block a page read for an order of magnitude longer time than read latency when it is no longer possible to delay the erase operation under heavy user writes.

\subsection{Incremental Step Pulse Erasure (\ispe)}
\label{ssec:motiv_ispe}
The lifetime and performance impact of erase operations increases even further in modern SSDs, as an erase operation often requires \emph{multiple erase loops.}
As a flash cell experiences more P/E cycles, the cell becomes more difficult to erase~\cite{hong-fast-2022,kim-pe-2021, micheloni2010inside}.
Consequently, an erase operation may fail to sufficiently erase all the cells in a target block, which we call an \emph{erase failure}.
To secure data reliability, it is common practice to employ the \underline{I}ncremental \underline{S}tep \underline{P}ulse \underline{E}rasure (ISPE) scheme~\cite{ispe, hollmer1999erase, aritome2015nand} that retries to erase the block with an increased erase voltage until completely erasing the block.

\begin{figure}[t]
\includegraphics[width=\linewidth]{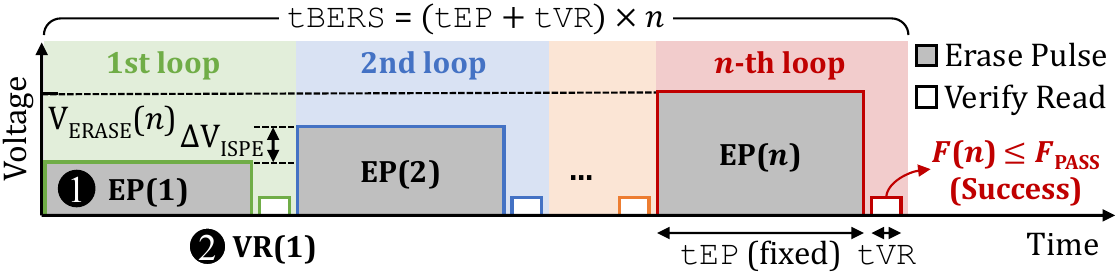}
\centering
\caption{Incremental Step Pulse Erasure (ISPE) scheme.}
\label{fig:ispe}
\end{figure}

\fig{\ref{fig:ispe}} shows how a NAND flash chip erases a block via multiple erase loops, each of which consists of two steps: \inum{i}~an erase-pulse (\ep) step and \inum {ii}~a verify-read (\vr) step.
An EP step (e.g., \bcirc{1}~in \fig{\ref{fig:ispe}}) applies \verase{} to the target block for a fixed amount of time \tep (e.g., 3.5 ms) that is predefined by NAND manufacturers at design time.
After each EP step, a VR step (e.g., \bcirc{2}) checks if all the cells in the block are sufficiently erased.
When \epi{i} (the $i$-th EP step, $i\geq{1}$) fails to do so, the ISPE scheme performs \epi{i+1} while progressively increasing \verase{} by a fixed amount \dverase. 
The ISPE scheme repeats this until completely erasing the block, leading to erasure latency \tbers as follows:
\begin{equation}
\tbers = (\tep + \tvr) \times N_\text{ISPE},
\end{equation}
\noindent
where \tvr is the VR latency ($\sim$100~\usec), and \nloop is the number of erase loops required to completely erase the block.

\begin{figure}[h]
\includegraphics[width=\linewidth]{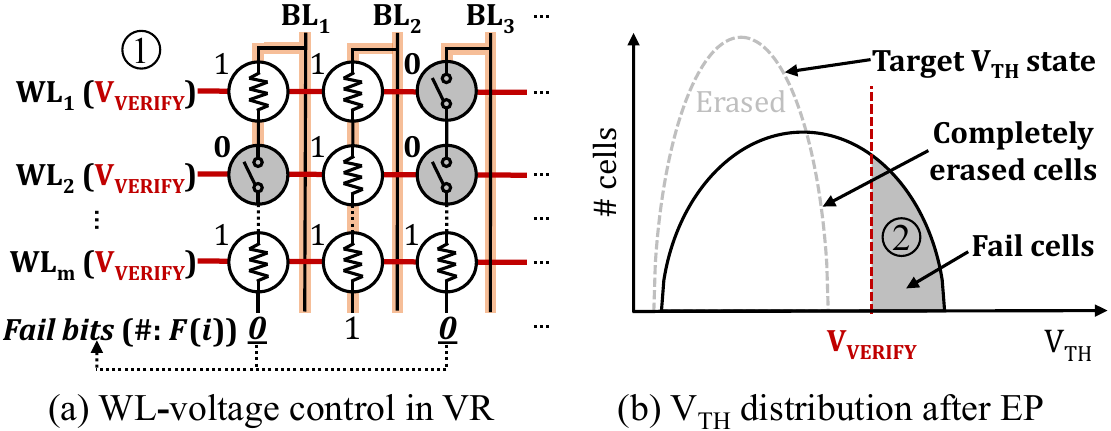}
\centering
\caption{Details of verify-read (VR) step in ISPE scheme.}
\label{fig:vr}
\end{figure}

\figs{\ref{fig:vr}a and \ref{fig:vr}b} describe how a NAND flash chip performs a \vr step.
It simultaneously senses (or activates) all the WLs in the block using a verify voltage \vverify (\wcirc{1} in \fig{\ref{fig:vr}a}) that is between the erase state and the first program state.
If \epi{i} fails, it means that the target block has some cells whose \vth levels are still higher than \vverify (\wcirc{2} in \fig{\ref{fig:vr}b}).
Such cells would operate as an off-switch during \vri{i} (the $i$-th \vr step) because $\text{V}_\text{TH}>\text{V}_\text{VERIFY}$, making the corresponding BLs read as `\texttt{0}' bits, called \emph{fail bits}.
Then, \vri{i} counts \nfail{i}, the number of fail bits after \epi{i}, using \emph{on-chip} counter logic~\cite{micheloni2010inside, ito2018sonos}.
It judges that \epi{i} succeeds only when \nfail{i} is lower than a predefined threshold \fpass.

\subsection{Limitations of Existing Techniques}
\label{sec:motivation:limits}
A large body of prior work~\cite{lee-ispass-2011,cui-date-2018,kang-dac-2018,shahidi-sc-2016,choi-hpdc-2018,guo-ipdps-2017,kang-cm-2017,pan-hotstorage-2019,lee-tcad-2013, murugan-msst-2011,li-msst-2019,dh-tcad-2022,jeong-hotstorage-2013, jeong-fast-2014, jeong-tc-2017, wu-fast-2012,kim-fast-2019, ispe} has proposed various optimizations to mitigate the negative impact of erase operations, but \emph{none} has yet investigated \emph{how erase latency should be set} to fully exploit the potential of NAND flash memory.
Like other memory technologies (e.g., DRAM), modern NAND flash memory also exhibits high process variation that leads to significantly different physical characteristics across flash cells.
Prior works have demonstrated varying physical characteristics across WLs~\cite{li-micro-2020, shim-micro-2019, hong-fast-2022, wang-acm-2017, chen-dac-2017, hung-ssc-2015, ha-tcad-2016} and blocks~\cite{kim-pe-2021, yen-hpca-2022, ha-tcad-2016, cai-dsn-2015} in modern 3D NAND flash memory.
Unfortunately, most existing techniques adopt the ISPE scheme that \emph{always} uses the \emph{same} erase-timing parameters (e.g., \tep) for \emph{every} block.
Doing so potentially causes unnecessary erase-induced cell stress, e.g., when a block can be erased more easily compared to the worst-case block.

\begin{figure}[t]
    \centering
    \includegraphics[width=\linewidth]{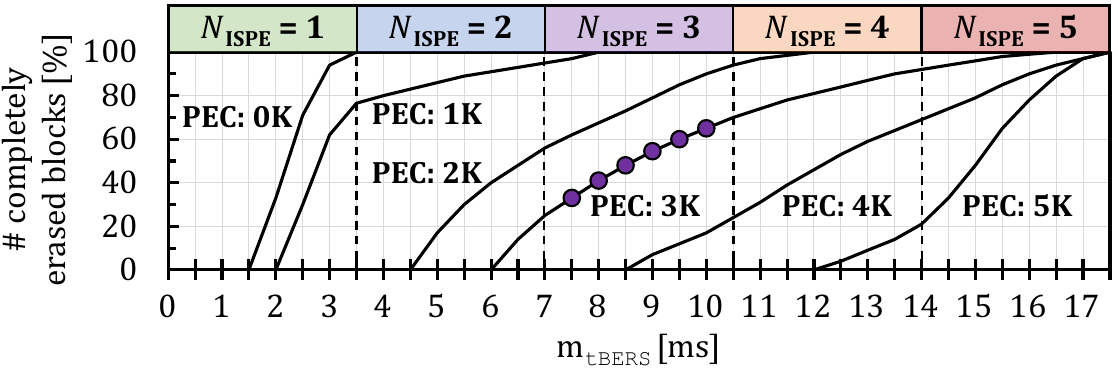}
    \caption{Erase latency variation under different P/E cycles.}
    \label{graph:erase_latency_per_pec}
\end{figure}

\head{Limitations of the ISPE Scheme}
To understand the potential of optimizing erase latency, we characterize 160 real 3D 48-layer triple-level cell (TLC) NAND flash chips (see \sect{\ref{sec:device_characterization_study}} for our characterization methodology).
We measure \mtbers, the minimum \tbers to completely erase a block, for 19,200 blocks randomly selected from the 160 chips.
\fig{\ref{graph:erase_latency_per_pec}} shows the cumulative distribution function (CDF) of $m_\texttt{tBERS}$($PEC$) across the tested blocks under different P/E-cycle counts ($PEC$).

We make three key observations.
First, an erase operation requires a number of erase loops in modern SSDs, significantly increasing not only erased-induced cell stress but also erase latency. 
While all the tested blocks can be erased via a single erase loop at zero PEC, \emph{every} block requires multiple \mbox{(2$\sim$4)} loops after 2K PEC, causing \mbox{2$\sim$4$\times$} increases in \tbers when using the ISPE scheme.
Second, \mtbers significantly varies even across blocks that require the same \nispe.
This clearly shows that a significant number of blocks are \emph{over erased} under the \ispe scheme, suffering from more erase-induced stress than necessary.
If we used the \ispe scheme, 40\% of the blocks at 3K PEC would require \nispe$=3$ and thus experience the high erase voltage for 10.5~ms (purple dots in \fig{\ref{graph:erase_latency_per_pec}}), while their \mtbers values significantly vary.
Third, there is considerable erase latency variation when \nispe$=1$.
More than 70\% (30\%) of the blocks at zero PEC (1K PEC) require only 2.5~ms to be completely erased, which is 29\% lower than \tbers in the \ispe scheme (3.5~ms).

We draw two conclusions based on our observations.
First, modern NAND flash memory suffers from unnecessarily longer erase latency (and thus more erase-induced cell stress) than actually needed.
Second, if it is possible to accurately predict and use \mtbers for a block, doing so would significantly mitigate the wear-out and long tail latency problems.

\begin{figure}[t]
    \centering
    \includegraphics[width=\linewidth]{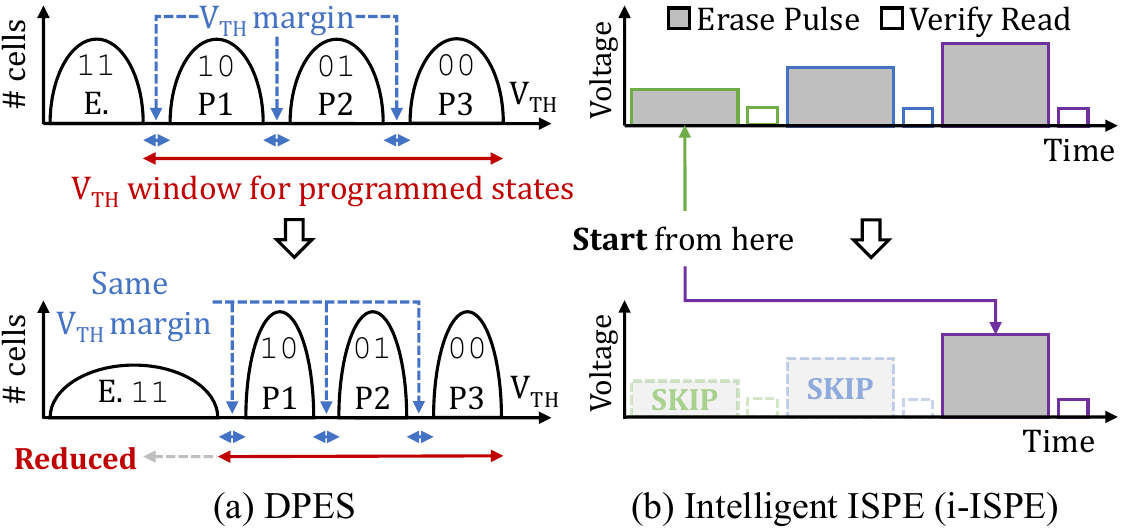}
    \caption{High-level overview of existing ISPE optimizations.}
    \label{fig:sota}
\end{figure}

\head{Limitations of the State-of-the-Art}
To our knowledge, only a few prior works on \emph{2D} NAND flash memory~\cite{jeong-hotstorage-2013, jeong-fast-2014, jeong-tc-2017, ispe} propose to dynamically tune ISPE parameters.
\fig{\ref{fig:sota}} depicts the high-level key ideas of (a) \emph{\dpes (\underline{D}ynamic \underline{P}rogram and \underline{E}rase \underline{S}caling)}~\cite{jeong-hotstorage-2013, jeong-fast-2014, jeong-tc-2017} and (b) \emph{\iispe (intelligent \ispe})~\cite{ispe}.
\dpes reduces erase-induced cell stress by decreasing \verase, which narrows the voltage window for the program states.
However, it also requires longer program latency to provide the same level of reliability as the original ISPE scheme by forming much narrower program \vth states.
I-ISPE tracks each block's \nispe to perform only the final erase loop \epi{N_\text{ISPE}} while skipping the previous loops (e.g., \epi{1} and \epi{2} in \fig{\ref{fig:sota}b}), which potentially reduces not only the erase-induced stress but also \tbers.

Unfortunately, it is challenging to adopt \dpes and \iispe in modern 3D NAND flash memory due to two reasons.
First, erasing 3D flash cells is more difficult compared to 2D cells due to the differences in cell physics and erase mechanisms between them~\cite{hong-fast-2022}.
Second, 3D NAND flash memory exhibits higher process variation across cells compared to 2D NAND flash memory~\cite{liu-asplos-2021,wang-acm-2017, chen-dac-2017}.
Such characteristics significantly limit the effectiveness of both \dpes and \iispe in modern NAND flash memory; \inum{i}~for \dpes, it becomes more challenging to secure the voltage window wide enough for the program states; \inum{ii}~for \iispe, skipping the first erase loops incurs an erase failure more likely, which, in turn, rather requires the next erase loop with a higher \verase (i.e., more erase-induced stress) compared to the conventional ISPE scheme.
We quantitatively evaluate the effectiveness of \dpes and \iispe in modern SSDs in \sect{\ref{sec:evaluation}}.

\section{\proposal: Adaptive Erase Operation}
\label{sec:key_idea}
In this work, we propose \emph{\proposal (\underline{A}daptive \underline{ER}ase \underline{O}peration)} that enhances SSD lifetime and improves I/O performance by applying \emph{near-optimal} erase latency for each target block.\linebreak
\textbf{The key idea} of \proposal is simple.
Unlike the \ispe scheme that performs all erase-pulse (\ep) steps with fixed latency \tep for every block, \proposal dynamically adjusts \tep to be just long enough for complete erasure of a target block.
\proposal uses the same \verasei{i} for each \epi{i} as the \ispe scheme, so \epi{i} in \proposal may also fail to completely erase the block even without \tep reduction.
In such a case, \proposal also performs the next \epi{i+1} while trying to reduce \tep \emph{if possible}, i.e., when it expects \epi{i+1} to completely erase the block with reduced \tep.
This means that \proposal reduces \tep in the \emph{final} erase loop (i.e.,~\epi{N_\text{ISPE}}) that completely erases the block, thereby reducing the total erase latency \tbers as follows:
\begin{equation}
\tbers = (\tep + \tvr ) \times N_\text{ISPE} - \Delta\tep,
\label{eq:tber_reduction}
\end{equation}
where $\Delta\tep$ is the amount of reduction in \tep in \epi{N_\text{ISPE}}. 

It would be ideal to dynamically optimize both \verase{} and \tep at the same time, but we decide to keep using the same \verase{} values in the \ispe scheme.
This is because it is quite challenging to accurately predict a near-optimal \verase{} for a block due to the high process variation in 3D NAND flash memory. 
Inaccurate \verase adjustment can cause insufficient or excessive erasure of flash cells as discussed in \sect{\ref{sec:motivation:limits}}, which can rather degrade SSD lifetime and I/O performance.
We leave accurate predictions and simultaneous optimizations of both \verase and \tep to future work.

\head{Fail-bit-count-based Erase Latency Prediction (\felp)} 
A key challenge in \proposal is to accurately identify \mtepi{i} for a block, i.e.,~the minimum value of \tep in each EP step (i.e.,~\epi{i}) just long enough to fully erase all the cells in the block.
Even though prior work~\cite{chen-trans-2018,lue-iedm-2015} has experimentally demonstrated a strong correlation between a block's PEC and erase latency, i.e., the higher the block's PEC, the longer the latency for erasing the block~\cite{jeong-tc-2017}, PEC alone is insufficient for accurate prediction of near-optimal \tep due to high process variation in modern NAND flash memory.
As shown in \fig{\ref{graph:erase_latency_per_pec}}, \mtep significantly varies even across blocks at the same PEC.
This suggests that \proposal needs a way more effective metric than PEC to figure out more precise erase characteristics (i.e., \mtepi{i}) for individual flash blocks.

\proposal addresses the challenge via \emph{\underline{F}ail-bit-count-based \underline{E}rase \underline{L}atency \underline{P}rediction (\felp)} that predicts \mtepi{i+1} based on \nfail{i}, the number of fail bits incurred by the previous \epi{i}.
Our key intuition is that \nfail{i} can likely be an \emph{accurate proxy} of \mtepi{i+1} because the more sufficiently the cells are erased by an EP step, the lower the fail-bit count.
Commodity NAND flash chips already calculate \nfail{i} for the ISPE scheme as explained in~\sect{\ref{ssec:motiv_ispe}}, so the implementation overhead of \felp is trivial (see \sect{\ref{sec:implementation}} for more detailed overhead analysis).

\begin{figure}[b]
    \centering
    \includegraphics[width=\linewidth]{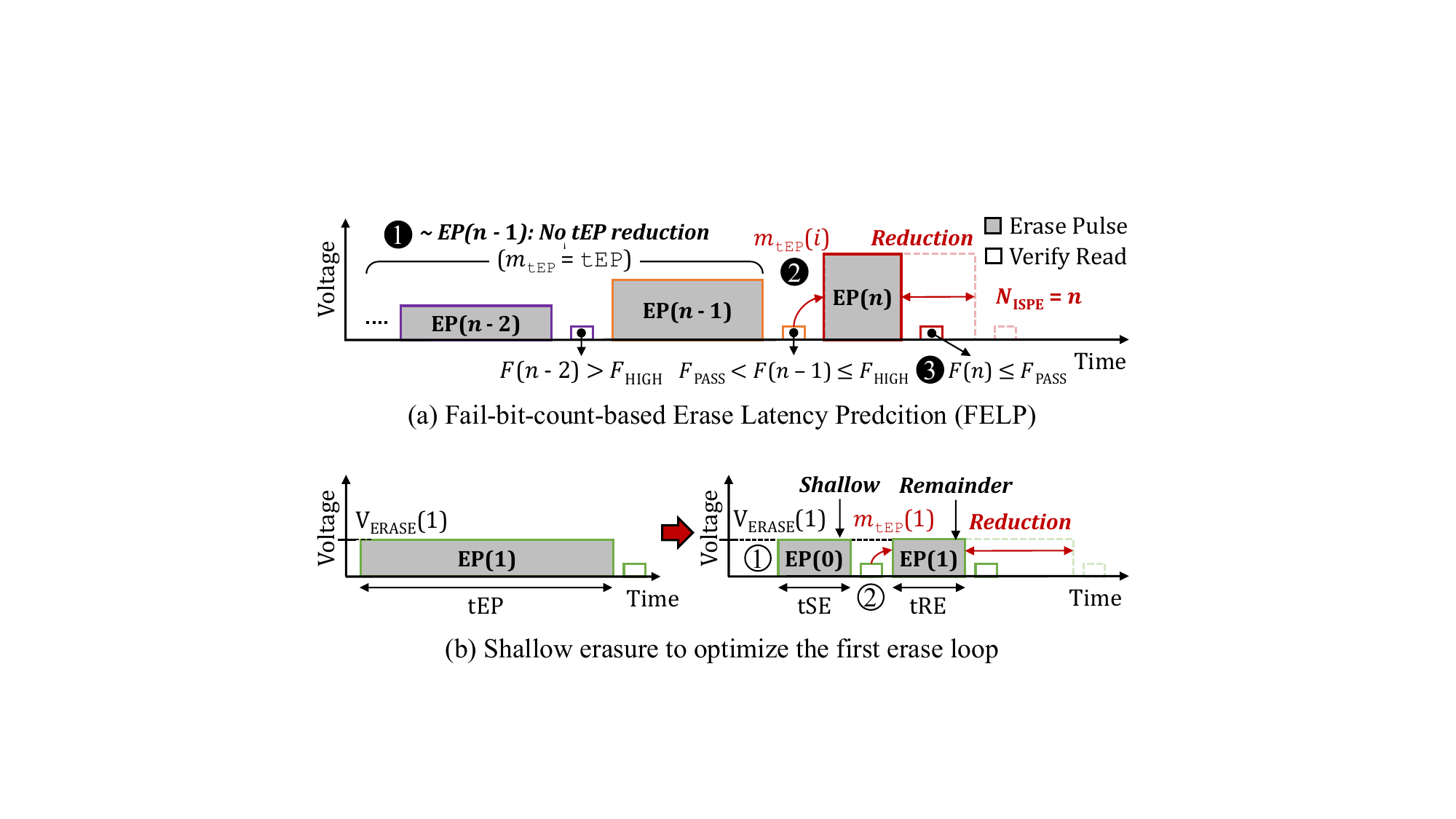}
    \caption{Erase latency reduction in \proposal.}
    \label{fig:felp}
\end{figure}

\fig{\ref{fig:felp}a} depicts how \proposal safely reduces \tep based on \felp.
Like the \ispe scheme, \proposal also performs a verify-read step \vri{i} after each \epi{i}, which results in \nfail{i}.
If \nfail{i} is higher than a threshold \fhigh ($\gg{F_\text{PASS}}$), \proposal uses the default \tep for the next \epi{i+1} considering that there is no room for \tep reduction (e.g., \bcirc{1} until \epi{N_\text{ISPE}-1} in \fig{\ref{fig:felp}a}).
When $F_\text{PASS}<F\text{(}i\text{)}\leq{F_\text{HIGH}}$,
\proposal \bcirc{2}~reduces \tep, i.e.,~it predicts and uses \mtepi{i+1} for \epi{i+1}, such that the lower the value of \nfail{i}, the lower the value of \mtepi{\nolinebreak{i+1}}.
Note that \proposal provides effectively the same reliability as the ISPE scheme as long as $F\text{(}N_\text{ISPE}\text{)}\leq{F_\text{PASS}}$ (\bcirc{3}).

\head{Shallow Erasure}
As \felp needs \nfail{i} to predict \mbox{\mtepi{i+1}}, it is difficult for \proposal to reduce \tep for the \emph{first} EP step, i.e.,~\epi{1}.
Optimizing \epi{1} is also important because a block can be completely erased via a single erase loop under many operating conditions.
As shown in \fig{\ref{graph:erase_latency_per_pec}}, 76.5\% of the tested blocks require only \epi{1} for complete erasure at 1K PEC (i.e., 20\% of the maximum PEC).

We enable \proposal to optimize \epi{1} by performing it in two steps.
\fig{\ref{fig:felp}b} describes how \proposal reduces the effective \tep for a single-loop erase operation.
Instead of using the default \tep (e.g., 3.5~ms) for \epi{1}, \proposal first performs a \emph{short} EP step with the default \verasei{1}, which we call \emph{shallow erasure}, for a reduced amount of time \tshallow (e.g., \wcirc{1} in \fig{\ref{fig:felp}b}).
It then \wcirc{2}~performs an additional VR step to obtain the number of fail bits after the shallow erasure (we denote shallow erasure as the \emph{0-th} erase loop in the rest of this paper, i.e.,~\epi{0}, \vri{0}, and \nfail{0} refer to the erase-pulse step, verify-read step, and fail-bit count in shallow erasure, respectively).
Finally, \proposal performs \emph{remainder erasure} that also applies the default \verasei{1} but dynamically adjusts the latency \texttt{tRE} (e.g., $0\leq{}\texttt{tRE}\leq{}2.5$~ms) based on \nfail{0}.

\head{Leveraging ECC-Capability Margin}
Prior work~\cite{shim-micro-2019, park-asplos-2021} has demonstrated that a large ECC-capability margin\footnote{The difference between the maximum number of bit errors per codeword that given ECC can correct and the number of bit errors in a codeword.} exists in modern SSDs due to two reasons.
First, modern SSDs commonly employ strong ECC to cope with the high RBER of NAND flash memory in the worst-case operating conditions (e.g., 1-year retention time at 5K PEC).
Second, it is common practice to employ \emph{read-retry} in modern SSDs~\cite{park-asplos-2021} to ensure data reliability.
When a read page's RBER exceeds the ECC capability, read-retry \emph{repeats} reading of the page with adjusted \vref{} until it sufficiently lowers RBER, thereby leading to a large ECC-capability margin.

\proposal leverages the large ECC-capability margin to reduce \tep \emph{more aggressively} by increasing the pass threshold \fpass in the ISPE scheme.
Doing so would likely cause incomplete erasure of a target block, which introduces additional bit errors.
However, we hypothesize that \proposal can still ensure data reliability in many cases due to three key reasons.
First, a block's reliability degrades as it experiences P/E cycling, so an even larger ECC-capability margin to tolerate the additional errors exists at low PEC.
Second, aggressive \tep reduction mitigates erase-induced cell stress, which can compensate for the additional errors as a long-term impact.
Third, a majority of additional fail cells due to increased $F_\text{PASS}$ would be programmed to higher \vth states under the data randomization technique~\cite{cha2013data, favalli2021scalable}, e.g., 87.5\% in TLC NAND flash memory.
It means that such fail cells are unlikely to cause bit error, significantly decreasing the reliability impact of aggressive \tep reduction.
To maximize \proposal's benefits without compromising data reliability, we enhance \felp to also consider the expected ECC-capability margin for a target block by keeping the number of additional errors caused by aggressive \tep reduction below the current ECC-capability.
\section{Device Characterization Study}\label{sec:device_characterization_study}
To validate our key ideas and hypothesis in \sect{\ref{sec:key_idea}}, we conduct an extensive real-device characterization study.

\subsection{Characterization Methodology}\label{sec:dev:method}
\head{Infrastructure} 
We use an FPGA-based testing platform with a custom flash controller and temperature controller.
The flash controller can perform not only basic NAND-flash commands (i.e., for read, program, and erase operations) but also low-level test commands such as \texttt{GET/SET FEATURE} commands~\cite{ONFI}.
This feature allows us to modify \tep of each \epi{i} at a granularity of 0.5~ms and obtain \nfail{i} from the chip after \vri{i}.
The temperature controller can maintain the operating temperature of the tested chips within \textpm\degreec{1} of the target temperature, thus minimizing unintended RBER variations potentially caused due to unstable temperature.

\head{Methodology} 
We characterize 160 real 48-layer 3D TLC NAND flash chips \rev{from Samsung~\cite{kang-isscc-2016}} in which the default $\texttt{tEP}=3.5$~ms.
To minimize the potential distortions in our results, for each test scenario, we evenly select 120 blocks from each chip at different physical locations and test all WLs in each selected block.
We test a total of 3,686,400 WLs (11,059,200 pages) to obtain statistically significant results.

We test the chips while varying PEC and retention time.
Unless specified otherwise, we increase a block's PEC by programming every page in the block using a random pattern and erasing the block with the default \tep in every erase loop.
We follow the JEDEC industry standard~\cite{jedec} for an accelerated lifetime test to analyze the reliability under the worst-case operating conditions.
For example, to emulate a 1-year retention time at \degreec{30}, we bake the chips at \degreec{85} for 13 hours following the Arrhenius’ law~\cite{arr}.

To identify a block's \mtbers (i.e., \nloop and \mtepi{N_\text{ISPE}}), we erase the block using a \emph{modified} ISPE scheme (m-ISPE) that we design by modifying the original ISPE scheme in two aspects.
First, we reduce the fixed latency \tep for each \epi{i} from 3.5~ms to 0.5~ms, i.e., we split an erase loop in the ISPE scheme into seven shorter loops.
Second, we increase \verase for every \emph{seven} erase loops (not for every loop) to effectively emulate the ISPE scheme.
If a block requires $n$ loops under \mbox{m-ISPE}, we estimate $N_\text{ISPE}=\lceil{n/7}\rceil$ and $m_\texttt{tEP}\text{(}N_\text{ISPE}\text{)}=0.5\times\text{(}1+\text{((}n - 1\text{)}~\text{mod}~7\text{))}$
of the block under the ISPE scheme.
\rev{Even though the m-ISPE scheme requires six additional ramping-up/down steps to charge/discharge voltage and VR steps for each erase loop compared to the original ISPE scheme, its reliability impact is negligible;
for our 160 tested chips, the m-ISPE scheme hardly increases the average RBER (by less than 1\%) compared to the original ISPE scheme, under 1-year retention time at 5K PEC.}

\subsection{Fail-Bit Count vs Near-Optimal Erase Latency}\label{sec:dev:fail}
\begin{figure}[h]
    \centering
    \includegraphics[width=\linewidth]{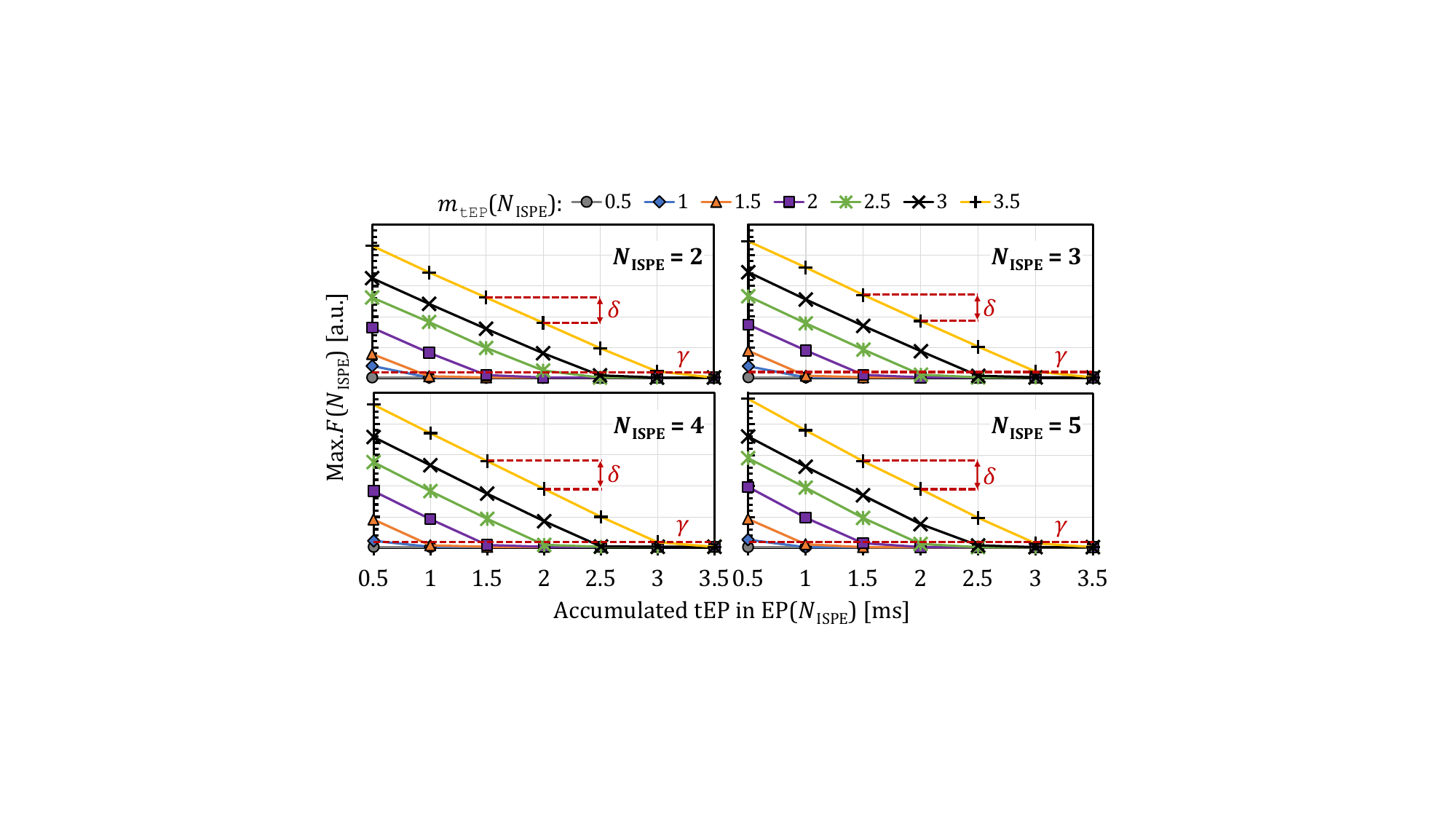}
    \caption{Impact of erase latency on the fail-bit count.}
    \label{graph:acc_tep}
\end{figure}

To validate the feasibility of \felp, we analyze the relationship between the 
minimum erase latency and fail-bit count.
We measure each block's \nloop and \mtep while tracking \nfail{i} in every \epi{i} under the m-ISPE scheme.
\fig{\ref{graph:acc_tep}} depicts the maximum value of \nfail{N_\text{ISPE}} within the blocks that have the same \mtep, when we progressively increase \tep by 0.5~ms in the final EP step (i.e.,~\epi{N_\text{ISPE}}).

We make two key observations.
First, the fail-bit count \emph{almost linearly} decreases as \tep increases.
While the negative correlation between \nfail{i} and \tep is expected in \sect{\ref{sec:key_idea}}, the correlation is significantly high and consistent; increasing \tep by 0.5~ms decreases \nfail{N_\text{ISPE}} by almost the same amount $\delta$ ($\simeq\text{5,000}$) in all tested blocks with different \nloop and \mtep values.
This suggests that \inum{i}~erase latency has a linear impact on the degree of erasure under the same erase voltage, and \inum{ii}~NAND manufacturers carefully set \verase values (i.e., \verasei{i} and \dverase) in the ISPE scheme to avoid excessive increases in erase-induced cell stress and erase latency. 
Second, when $m_\texttt{tEP}\text{(}N_\text{ISPE}\text{)}=0.5$~ms, \nfail{N_\text{ISPE}} is quite consistent at a certain value $\gamma$ ($\ll\delta$) in all test scenarios.
The result suggests that the lower the cell's \vth level, the more difficult it becomes to further reduce the \vth level.

\begin{figure}[h]
    \centering
    \includegraphics[width=\linewidth]{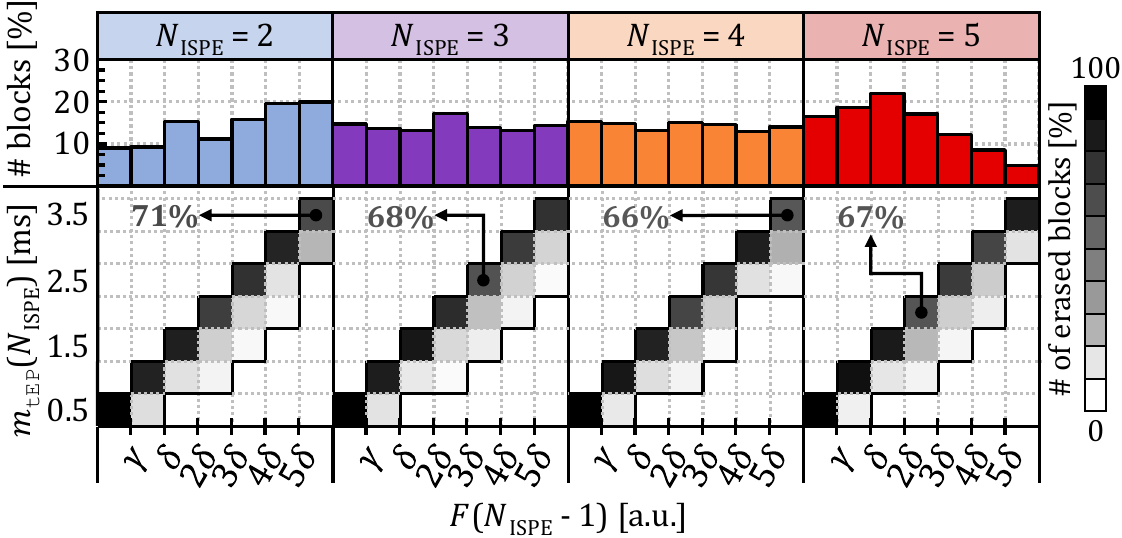}
    \caption{Erase-pulse latency depending on the fail-bit count.}
    \label{graph:failbits_vs_tbers}
\end{figure}

Our observations highlight the high potential of \felp.
To confirm this, we analyze how a block's \mtep varies depending on its \nfail{N_\text{ISPE}-1}.
\fig{\ref{graph:failbits_vs_tbers}} depicts the probability of \mtep at different $N_\text{ISPE}$ across the fail-bit ranges that we set based on $\gamma$ and $\delta$ from \fig{\ref{graph:acc_tep}}. 
A box at \mbox{($x$, $y$)} in \fig{\ref{graph:failbits_vs_tbers}} represents the probability (in grayscale) that a block requires $m_\texttt{tEP}\text{(}N_\text{ISPE}\text{)}=y$~ms for complete erasure when its \mbox{\nfail{N_\text{ISPE}-1}} belongs to $x$-th fail-bit range.
We also plot the fraction of blocks that belong to the $x$-th fail-bit range (top).
Note that for the same $N_\text{ISPE}$, the sum of all fractions (top) is 100\%, and the sum of all probabilities at the same $x$-th fail-bit range (bottom) is 100\%.

We make two key observations.
First, \felp is highly effective at predicting \mtep.
A majority of blocks (e.g., $\geq66\%$ in $N_\text{ISPE}=4$) in the same fail-bit range require the same \mtep under all different $N_\text{ISPE}$ cases.
Even though every fail-bit range contains some blocks whose \mtep is lower compared to the majority of blocks, the fraction is quite low (e.g., $<34\%$ in $N_\text{ISPE}=4$) in all the fail-bit ranges and \nloop cases.
Second, \nfail{N_\text{ISPE}-1} is distributed across blocks quite evenly in all \nloop cases.
This highlights again that \nfail{N_\text{ISPE}-1} is an accurate proxy of \mtep, given that \mtep also significantly varies across blocks in a wide range as shown in \fig{\ref{graph:erase_latency_per_pec}}.

Based on our observations, we draw two conclusions.
First, \proposal can accurately predict the minimum erase latency using FELP, even for blocks that have varying erase characteristics.
Second, the implementation of FELP requires only identifying two values for fail-bit ranges, e.g., $\gamma$ and $\delta$ in \figs{\ref{graph:acc_tep} and \ref{graph:failbits_vs_tbers}}, which can be done using our characterization methodology.

\subsection{Shallow Erasure: Feasibility \& Parameter Setting}\label{sec:dev:shallow}
\begin{figure}[h]
    \centering
    \includegraphics[width=\linewidth]{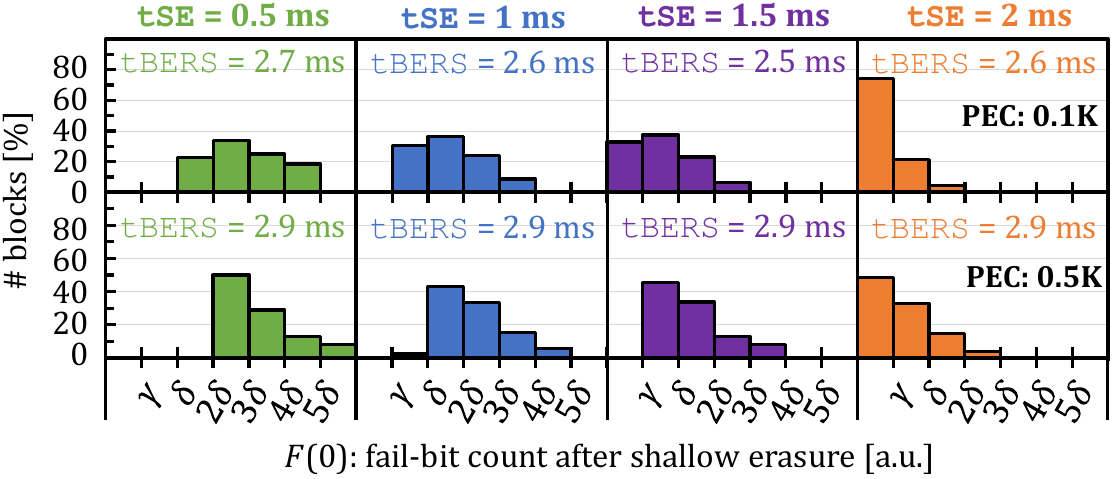}
    \caption{Fail-bit distribution under varying \tshallow.}
    \label{graph:shallow_erase}
\end{figure}

We validate the feasibility of shallow erasure and explore a good latency value.
\fig{\ref{graph:shallow_erase}} shows the distribution of \nfail{0} across blocks
under different $\texttt{tSE}=\langle$0.5~ms, 1~ms, 1.5~ms, 2~ms$\rangle$ and $PEC=\langle$0.1K, 0.5K$\rangle$.
We confirm that all the tested blocks in the $n$-th fail-bit range can be completely erased via subsequent remainder erasure with $\texttt{tRE}=0.5\times{n}$~ms.

We make three key observations from \fig{\ref{graph:shallow_erase}}.
First, shallow erasure enables \proposal to reduce \tep for \epi{1}.
A~large fraction of blocks can be completely erased with lower latency than the default \tep ~($\langle$80\%, 85\%, 86\%, 88\%$\rangle$ for $\texttt{tSE}=\langle$0.5~ms, 1~ms, 1.5~ms, 2~ms$\rangle$).
Second, \tshallow does \emph{not} significantly affect the average \tbers.
We annotate the average \tbers for each \tshallow case in \fig{\ref{graph:shallow_erase}}, which shows their small variation ($<$10\%).
Third, with a low probability, shallow erasure may cause a block to be over-erased.
For example, at 0.5K PEC, 2\% blocks can be erased with \tbers$=2$~ms when \tshallow$=$1~ms (the second fail-bit range), while there exist no such blocks in the other \tshallow cases.

Based on our observations, we conclude that \proposal can significantly optimize not only multi-loop erase operations but also single-loop erase operations using shallow erasure.
For our tested chips, we decide to set \tshallow$=1$~ms to minimize the probability of an over-erased block, which reduces both erased-induced stress for 85\% of blocks and the average erase latency by 21\% for single-loop erase operations.

\subsection{Reliability Margin for Aggressive \tep Reduction}
\label{sec:dev:aggressive}
\begin{figure}[h]
    \centering
    \includegraphics[width=\linewidth]{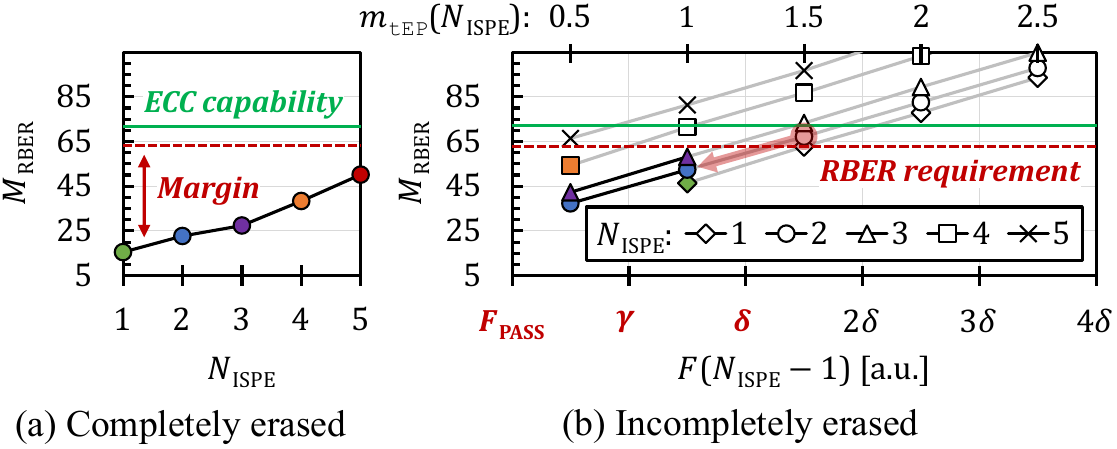}
    \caption{Reliability margin depending on erase status.}
    \label{graph:rber_opt}
\end{figure}

To identify the ECC-capability margin for more aggressive \tep reduction, we analyze the reliability impact of insufficient erasure.
We measure \mrber of each block, i.e., the maximum RBER within the pages in the block under 1-year retention time at \degreec{30}, when we erase the block in two different ways. 
First, we \emph{completely} erase the block by performing \nloop erase loops with \mtep.
Second, we \emph{insufficiently} erase the block by performing only ($N_\text{ISPE}-1$) erase loops, which results in $F_\text{PASS}<F\text{(}N_\text{ISPE}-1\text{)}\leq{F_\text{HIGH}}$.
\fig{\ref{graph:rber_opt}} depicts the maximum value of \mrber within the tested blocks when we program the blocks (a) after complete erasure and (b) after insufficient erasure.
For \fig{\ref{graph:rber_opt}b}, we group the tested blocks depending on their \nloop and fail-bit range.
We also plot \inum{i}~the \emph{ECC capability} at 72 bits per 1 KiB and \inum{ii}~the \emph{RBER requirement} at 63 bits per 1 KiB to reflect sampling errors (i.e., a block is considered unusable if its $M_\text{RBER}>63$ to incorporate a safety margin into the ECC capability).

We make two key observations.
First, we observe a large reliability margin (i.e., $RBER~requirement-M_\text{RBER}$) that can potentially be used to further reduce \tep especially in the early lifetime stage of blocks.
As shown in \fig{\ref{graph:rber_opt}a}, when a block is completely erased, there always exists a positive reliability margin for all \nloop values up to 47 bit errors ($N_\text{ISPE}=1$).
Second, it is possible to further reduce \tep without compromising reliability in many operating conditions.
As shown in \fig{\ref{graph:rber_opt}b}, using an insufficiently-erased block still meets the RBER requirement (i.e., \mbox{$M_\text{RBER}<63$}) if either of the following two conditions is met: \textbf{[C1:}~$N_\text{ISPE}\leq3$ and $F\text{(}N_\text{ISPE}-1\text{)}<\delta$\textbf{]} and \textbf{[C2:}~$N_\text{ISPE}=4$ and $F\text{(}3\text{)}<\gamma$\textbf{]}.
This means that we can skip the final erase loop in such cases, thereby further increasing the amount of \tep reduction even higher than the default \tep in the ISPE scheme, i.e., $\Delta\texttt{tEP}$ can be larger than \tep in \eqref{eq:tber_reduction}.
Note that \proposal can further reduce \mtep even if neither of \textbf{[C1]} and \textbf{[C2]} is met because increasing \tep by 0.5~ms in \epi{i} decreases \nfail{i} by $\delta$ as demonstrated in \sect{\ref{sec:dev:fail}}.
For example, a block requires $m_\texttt{tEP}\text{(}N_\text{ISPE}\text{)}=1.5$~ms for complete erasure when $N_\texttt{ISPE}=3$ and $\delta<F\text{(}2\text{)}\leq2\delta$, but using $\texttt{tEP}=0.5$~ms in \epi{3} can still meet the RBER requirement since doing so would decrease \mrber below 63 (see the arrow in \fig{\ref{graph:rber_opt}}b).

We conclude that \proposal can significantly mitigate the negative impact of erase operations by leveraging not only the high process variation but also the large ECC-capability margin present in modern SSDs.
Table~{\ref{tab:mtep}} shows the final \mtep model that we construct;
each cell's value `$t_1$~/~$t_2$' indicates \mtep when \proposal leverages only the process variation ($t_1$) and when also leveraging the ECC-capability margin ($t_2$). 

\begingroup
\def\arraystretch{1.1}
\newcommand\hmg{9pt}
\begin{table}[h]
\renewcommand\theadfont{\bfseries}
\caption{Final model of \mtepi{N_\text{ISPE}} based on \nfail{N_\text{ISPE}-1}.}
\label{tab:mtep}
\centering
\resizebox{\columnwidth}{!}{%
\begin{tabular}{@{\hspace{0pt}}c@{\hspace{5pt}}c@{\hspace{\hmg}}c@{\hspace{\hmg}}c@{\hspace{\hmg}}c@{\hspace{\hmg}}c@{\hspace{\hmg}}c@{\hspace{\hmg}}c@{\hspace{\hmg}}c@{\hspace{0pt}}}
\toprule
\multirowcell{2}{\textbf{\nloop}} & \multicolumn{8}{c}{\textbf{\nfail{N_\text{ISPE}-1}}} \\\cmidrule{2-9}
& $\leq\gamma$ & $\leq\delta$ & $\leq2\delta$ & $\leq3\delta$ & $\leq4\delta$ & $\leq5\delta$ & $\leq6\delta$ & $\leq7\delta$ \\
\midrule
\textbf{1} & 0.5 / \textbf{0} & 1 / \textbf{0} & 1.5 / \textbf{0.5} & 2 / \textbf{1} & 2.5 / \textbf{1.5} & 2.5 / \textbf{2} & 2.5 / 2.5 & 2.5 / 2.5 \\\cmidrule{2-9}
\textbf{2} & 0.5 / \textbf{0} & 1 / \textbf{0} & 1.5 / \textbf{0.5} & 2 / \textbf{1} & 2.5 / \textbf{1.5} & 3 / \textbf{2} & 3.5 / \textbf{2.5} & 3.5 / \textbf{3} \\\cmidrule{2-9}
\textbf{3} & 0.5 / \textbf{0} & 1 / \textbf{0} & 1.5 / \textbf{0.5} & 2 / \textbf{1} & 2.5 / \textbf{1.5} & 3 / \textbf{2} & 3.5 / \textbf{2.5} & 3.5 / \textbf{3} \\\cmidrule{2-9}
\textbf{4} & 0.5 / \textbf{0} & 1 / \textbf{0.5} & 1.5 / \textbf{1} & 2 / \textbf{1.5} & 2.5 / \textbf{2} & 3 / \textbf{2.5} & 3.5 / \textbf{3} & 3.5 / 3.5 \\\cmidrule{2-9}
\textbf{5} & 0.5 / 0.5 & 1 / 1 & 1.5 / 1.5 & 2 / 2 & 2.5 / 2.5 & 3 / 3 & 3.5 / 3.5 & 3.5 / 3.5 \\
\bottomrule
\end{tabular}}
\end{table}
\endgroup

\subsection{\rev{Applicability of \proposal for Other Types of Chips}}\label{sec:dev:applicability}

We expect that the key ideas of \proposal are generally applicable to a wide range of NAND flash chips other than the chips used for our characterization study due to three reasons.
First, our chips well represent modern 3D NAND flash memory because most commercial chips including SMArT/TCAT/BiCS have similar structures and cell types, e.g., vertical channel structures, gate-all-around cell transistors, and charge-trap type flash cells~\cite{micheloni2010inside, micheloni3d, aritome2015nand}, sharing key device characteristics like operation mechanisms and reliability characteristics.
Second, the erase mechanism of NAND flash memory has not changed significantly for more than a decade. 
For example, the ISPE scheme has been used since 2D SLC NAND flash memory~\cite{micheloni2010inside, aritome2015nand}.
Third, AERO does \emph{not} rely on chip-specific behaviors but leverages inherent erase characteristics, e.g., the more completely the cells within a block are erased, the lower the fail-bit count of the block.

\begin{figure}[h]
    \centering
    \includegraphics[width=\linewidth]{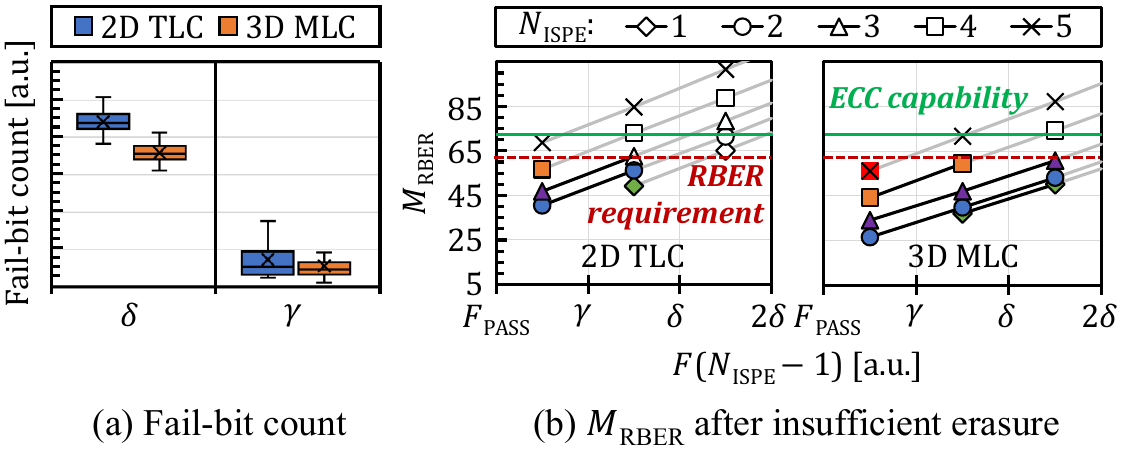}
    \caption{\rev{Erase characteristics of other chip types.}}
    \label{graph:chip_summary}
\end{figure}

To support our hypothesis, we characterize two additional types of NAND flash chips, \inum{i} 2x-nm 2D TLC NAND flash chips~\cite{vatto-anandtech-2012} and \inum{ii} 48-layer 3D MLC NAND flash chips~\cite{tallis-anandtech-2012} from Samsung.
We use the same methodology as other device characterizations.
\fig{\ref{graph:chip_summary}} shows (a) the values of $\delta$ and $\gamma$ for all \tep and \nispe cases (box plot) and (b) the maximum value of \mrber within the tested block after insufficient erasure.

We make two key observations.
First, although the exact values of $\delta$ and $\gamma$ slightly vary depending on the chip type, they are quite consistent within the same type of chips across all tested cases as shown in \fig{\ref{graph:chip_summary}a}.
This clearly shows that the strong linear relationship between the fail-bit count and accumulate \tep also holds in the chips additionally tested.
Second, as shown in \fig{\ref{graph:chip_summary}b}, the reliability impacts of insufficient erasure in the 2D TLC and 3D MLC chips exhibit highly similar trends to those in the 3D TLC chips (cf. \fig{\ref{graph:rber_opt}b}),\footnote{The results shown in \figs{\ref{graph:rber_opt}b and \ref{graph:chip_summary}b}, which are remarkably similar, might seem counter-intuitive because it is well known that 3D NAND flash memory exhibits better reliability compared to 2D NAND flash memory~\cite{luo-acm-2018}.
Nevertheless, commodity 3D NAND flash chips provide a similar level of reliability to 2D chips because the better reliability characteristics of 3D NAND flash memory are used to avoid write-performance degradation over 2D NAND flash memory.
Since its first development, a majority of 3D NAND flash memory has been designed to contain two planes~\cite{old_3d}, whereas state-of-the-art 2D chips consist of four planes~\cite{recent_2d}. The decrease in the number of planes per chip would inevitably limit the bandwidth of multi-plane writes in 3D NAND flash chips to be around twice lower compared to 2D chips if the same program-timing parameters were used (e.g., \tprog~$\approx$~1.2~ms in 2D NAND flash memory~\cite{recent_2d}). To avoid this, manufacturers have reduced the program latency of 3D NAND flash memory by trading its better reliability, which enables sustaining (or even improving) the write bandwidth of 3D chips without compromising reliability compared to 2D chips.
} suggesting the high feasibility of aggressive \tep reduction also in the two types of chips.
We conclude that \proposal can be used for a wide range of chips with the same methodology we use to construct the \tep model for our tested chips (Table~\ref{tab:mtep}).
\section{Design and Implementation}\label{sec:implementation}
We design \impl, an \proposal-enabled flash translation layer (FTL), by extending the conventional page-level FTL~\cite{gupta-2009-asplos} with two key data structures: \inum{i}~\emph{\underline{E}rase-timing \underline{P}arameter \underline{T}able (EPT)} and \inum{ii}~\emph{\underline{S}hallow \underline{E}rasure \underline{F}lags (SEF)}.
The EPT is a simple table to stores \mtepi{i} for each \epi{i} depending on \nfail{i-1}, which can be built via offline profiling of target chips as in~\sect{\ref{sec:device_characterization_study}} (Table~\ref{tab:mtep}).
The SEF is a bitmap that keeps track of whether a block needs shallow erasure or not.
Every bit in the SEF is initially set to `\texttt{0}', which is translated to \texttt{TRUE}. Doing so enables \impl to always perform shallow erasure for a fresh block that has experienced no P/E cycle.

\begin{figure}[h]
    \centering
    \includegraphics[width=\linewidth]{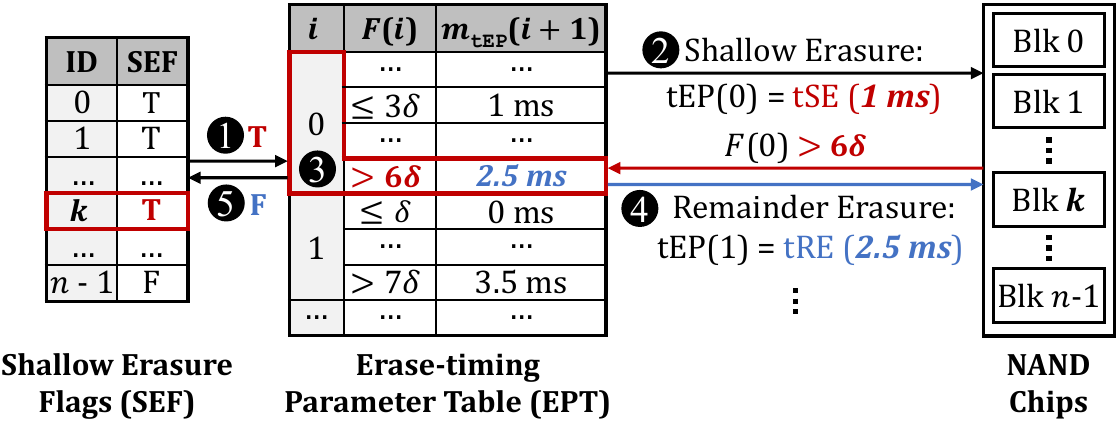}
    \caption{Operational overview of \impl.}
    \label{fig:impl}
\end{figure}

\fig{\ref{fig:impl}} illustrates how \impl dynamically adjusts erase latency.
For erasing a block whose index (or ID) is $k$, it first looks up the corresponding ($k$-th) bit in the SEF (\bcirc{1} in \fig{\ref{fig:impl}}).
If the flag bit is \texttt{TRUE}, \impl~\bcirc{2}~performs shallow erasure by changing the target chip's \tep to \tshallow with a \texttt{SET FEATURE} command.
Then it \bcirc{3}~queries the EPT with \nfail{0} that can be obtained via a \texttt{GET FEATURE} command.
Based on the query result, \impl \bcirc{4}~sets the chip's \tep and performs remainder erasure.
If the remainder erasure \emph{cannot} reduce the effective latency of the first erase loop, \impl~\bcirc{5} sets the corresponding bit in the SEF to `\texttt{1}' that is translated to `\texttt{FALSE}'.
Doing so allows \impl to skip shallow erasure for the block in future, but directly performs \epi{1} with the default \tep, thereby avoiding the unnecessary \vri{0}.
The remaining process is straightforward. 
If an erase loop fails, \impl performs the next loop while adjusting the chip's \tep based on the value obtained from the EPT. 

\head{Misprediction Handling}
While \emph{not} observed in our real-device characterization, 
it is possible that \proposal fails to completely erase outlier blocks (which could potentially be missed in the set of blocks we experimentally test) with reduced latency.
\impl can easily handle such a misprediction because it always checks \nfail{N_\text{ISPE}} anyways to verify whether a block is completely erased.
If \impl detects a misprediction, it repeats an additional EP step with appropriate \verase and \tep until completely erasing the block; it uses the same \verase if the accumulated latency is lower than the default \tbers while increasing \verase otherwise.
Despite \proposal's low misprediction rate, we evaluate its potential impact in \sect{\ref{sec:evaluation}}.

\head{Impact on ECC-Decoding Latency}
Aggressive \texttt{tEP} reduction in \proposal can potentially increase ECC-decoding latency (e.g., in LDPC~\cite{ldpc}), but its performance impact is limited due to two reasons.
First, it is common practice to limit the number of iterations for hard-decision ECC decoding~\cite{vahabzadeh-fms-2016, declercq-fms-2019} so that ECC-decoding latency (e.g., 8~\usec~\cite{ldpc_decision}) can be hidden by page sensing and data transfer.
Second, the RBER requirement conservatively set in \proposal (e.g., 63 bits per 1 KiB) ensures a low hard-decision failure rate (e.g., $<10^{-5}$~\cite{ldpc_decision}), hardly causing additional soft-decision LDPC decoding.

\head{Multi-Plane Operations}
Multi-plane operation is a widely used feature in modern NAND flash-based SSDs to improve I/O performance~\cite{micheloni2010inside,flash-cosmos,genstore}.
A typical NAND flash chip consists of multiple (e.g., 2 or 4) \emph{planes}, each of which has its own flash cell array and page buffer.
Planes in the same chips can operate concurrently if certain conditions are met, e.g., only the same type of NAND flash operations (i.e., read, program, or erase operations) can be performed in parallel across planes in a chip because they share part of peripheral circuitry.
Multi-plane operation can increase chip throughput linearly with the number of planes per chip, thereby significantly improving SSD-internal parallelism and I/O performance.

\proposal can work with multi-plane erase operations due to two reasons.
First, a NAND flash chip can individually set \tep{} for each target block of a multi-plane erase operation.
Second, as soon as a target block is completely erased, it is possible to \emph{inhibit} the block from being further erased by the subsequent erase loops during a multi-plane erase operation.
Although the worst block (which requires the longest erase latency) in target planes determines the latency of a multi-plane erase operation, \proposal can still sustain lifetime benefit and reduce tail latency by erasing each block in target planes only with necessary loops and times.

\head{Implementation Overhead}
\impl requires only two small changes to conventional SSDs.
First, \impl can use \texttt{GET/SET FEATURE} commands to obtain \nfail{i} and adjust \tep for each \epi{i}, respectively, thereby requiring no change to commodity NAND flash chips.
Second, the storage overhead for the EPT and SEF is trivial. 
The EPT needs to keep $T\times{L}$ entries, where $T$ and $L$ indicate the number of possible \tep values and the maximum number of erase loops, respectively.
In our current design, the EPT has 35 entries ($T=7$ and $L=5$), which requires only 140 bytes even when using a 32-bit value per entry.
The SEF needs to keep 1-bit information for each block. 
In our tested chips, the block size is around 10~MB, so the SEP's storage overhead is $1.25\times10^{-6}$\% ($=1/\text{(}8\times10^7\text{)}$) of SSD capacity, e.g., 12.5~KB for a 1-TB SSD.
Note that modern SSDs contain several GB of internal DRAM~\cite{ssd_dram}, so the storage overhead is negligible.
\section{Evaluation}\label{sec:evaluation}
We evaluate the effectiveness of \proposal at improving the lifetime and performance of modern NAND flash-based SSDs.
\subsection{Evaluation Methodology}\label{sec:eval:method}
We evaluate \proposal in two ways.
First, we characterize 160 real 3D TLC NAND flash chips to assess the lifetime enhancement of \proposal.
Unless specified otherwise, we follow the real-device characterization methodology explained in \sect{\ref{sec:dev:method}}.
Second, we evaluate the impact of \proposal on I/O performance using MQSim~\cite{tavakkol-fast-2018}, a state-of-the-art SSD simulator.

We compare five different erase schemes: \inum{i}~\basessd, \mbox{\inum{ii}~\iispessd}, \inum{iii}~\dpesssd, \inum{iv}~\aerossd, and \inum{v}~\aerooptssd.
\basessd is the conventional ISPE scheme explained in \sect{\ref{ssec:motiv_ispe}}.
\textsf{I-ISPE} is the intelligent ISPE scheme~\cite{ispe} explained in \sect{\ref{sec:motivation:limits}}, which directly performs \epi{n} while skipping the previous EP steps if the block has been completely erased by \epi{n} in the most-recent erase operation.
\dpesssd (explained in \sect{\ref{sec:motivation:limits}}) mitigates erase stress by reducing erase voltage \verase by 8--10\% at a cost of 10--30\% increase in write latency \tprog~\cite{jeong-tc-2017}.
Since \dpesssd is only applicable until 3K PEC in our tested chips (i.e., no matter how much \tprog increases, reducing \verase can no longer meet the reliability requirements), we use the same \verase and \tprog values as in \basessd after 3K PEC.
\aerossd and \aerooptssd dynamically adjust \tep for each \epi{i} at a granularity of 0.5~ms based on \nfail{i-1}.
\aerossd does not exploit the ECC-capability margin, i.e., it is more \emph{conservative} in \tep reduction compared to \aerooptssd that adopts all optimizations in \sect{\ref{sec:key_idea}}.

\head{Simulation Methodology}
We extend MQSim in two aspects to model the behavior of modern SSDs more faithfully.
First, we modify the NAND flash model of MQSim to emulate the erase characteristics of our 160 tested chips.
To this end, during our real-device characterization study in \sect{\ref{sec:device_characterization_study}}, we keep track of erase-related metadata such as the minimum erase latency, fail-bit count, and PEC for every tested block.
For simulation, we then randomly select tested blocks and assign their metadata to each of the simulated blocks in MQSim.
Because MQSim already tracks PEC, a simulated block can accurately emulate the erase characteristics of the corresponding real block at a given PEC by simply looking up the metadata.
Second, we optimize the request scheduling algorithm of MQSim to service user I/O requests with a higher priority over SSD-internal read/write/erase operations, e.g., suspending an ongoing erase operation~\cite{kim-fast-2019}.
Table~\ref{tab:sim_config} summarizes the configurations of the simulated SSDs.
We configure the architecture and timing parameters of simulated SSDs to be close to commodity high-end SSDs (e.g.,~\cite{samsung}).


\begingroup
\def\arraystretch{1.2}
\begin{table}[h]
\caption{Configurations of simulated SSDs.}
\label{tab:sim_config}
\centering
\resizebox{\columnwidth}{!}{
\begin{tabular}{ccc}
\toprule
\multirow{3}{*}{\textbf{SSD}} & Capacity: 1024 GB & Interface: PCIe 4.0 (4 lanes)\\
& GC policy: greedy~\cite{chang-2002-rtas} & Overprovisioning ratio: 20\% \\
& \# of channels: 8 & \# of chips per channel: 2 \\
\midrule
\multirowcell{6}{\textbf{NAND Flash} \\ \textbf{Chip}} & \# of planes per chip: 4 & \# of blocks per plane: 497 \\
& \# of pages per block: 2,112 & Page size: 16 KB \\
& MLC technology: TLC  & \texttt{tR}: 40 \usec~\cite{cho-isscc-2021} \\
& \tep (\proposal): 0.5~ms -- 3.5~ms & \tep: 3.5~ms~\cite{cho-isscc-2021} \\
& \tshallow (\proposal): 1~ms  & \texttt{tPROG}: 350 \usec~\cite{cho-isscc-2021} \\
& \multicolumn{2}{c}{\texttt{tPROG}: 385 \usec (\dpesssd, 0.5K PEC), 455 \usec (\dpesssd, 2.5K PEC)} \\
\bottomrule
\end{tabular}
}
\end{table}
\endgroup

For our performance evaluation, we study eleven workloads selected from two benchmark suits, Alibaba Cloud Traces~\cite{alibaba} and Microsoft Research Cambridge (MSRC) Traces~\cite{msr}, which are collected from real datacenter and enterprise servers.
For MSRC traces, we reduce the inter-request arrival time by 10$\times$, as similarly done in a large body of prior work to evaluate more realistic workloads~\cite{tavakkol-fast-2018, nadig-isca-2023, hong-fast-2022, yen-hpca-2022, mix_liu_1, mix_liu_2, mix_lv, mix_wu, jeong-tc-2017, jeong-fast-2014, jeong-hotstorage-2013}.
Table~\ref{tab:bench} summarizes the I/O characteristics of the workloads used for our evaluation.

\begingroup
\def\arraystretch{0.85}
\begin{table}[h]
\renewcommand\theadfont{\bfseries}
\caption{I/O characteristics of evaluated workloads.}
\label{tab:bench}
\centering
\resizebox{\columnwidth}{!}{
\begin{tabular}{cccccc}
\toprule
\thead{Benchmark} & \thead{Trace} & \thead{Abbr.} & \thead{Read \\ Ratio} & \thead{Avg. Req. \\ Size} & \thead{Avg. Inter Req. \\ Arrival Time}\\
\midrule
\multirowcell{6.5}{Alibaba \\ Cloud~\cite{alibaba}} & ali\_32 & ali.A & 7\% & 54~KB & 16.3~ms \\\cmidrule{2-6}
& ali\_3 & ali.B & 52\% & 26~KB & 111.8~ms \\\cmidrule{2-6}
& ali\_12 & ali.C & 69\% & 38~KB & 57.9~ms \\ \cmidrule{2-6}
& ali\_121 & ali.D & 78\% & 18~KB & 13.8~ms \\ \cmidrule{2-6}
& ali\_124 & ali.E & 95\% & 36~KB & 5.1~ms \\
\midrule
\multirowcell{8}{MSR \\ Cambridge~\cite{msr}} & rsrch\_0 & rsrch & 9\% & 9~KB & 421.9~ms \\\cmidrule{2-6}
& stg\_0 & stg & 15\% & 12~KB & 297.8~ms \\\cmidrule{2-6}
& hm\_0 & hm & 36\% & 8~KB & 151.5~ms \\\cmidrule{2-6}
& prxy\_1 & prxy & 65\% & 13~KB & 3.6~ms \\\cmidrule{2-6}
& proj\_2 & proj & 88\% & 42~KB & 20.6~ms \\\cmidrule{2-6}
& usr\_1 & usr & 91\% & 49~KB & 13.4~ms \\
\bottomrule
\end{tabular}}
\end{table}
\endgroup

\subsection{Impact on SSD Lifetime}\label{sec:eval:lifetime}
\begin{figure}[h]
    \centering
    \includegraphics[width=\linewidth]{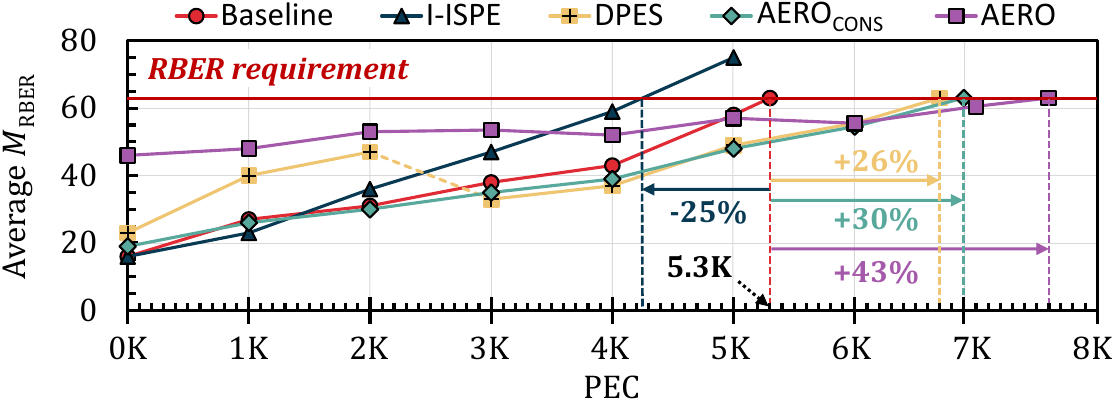}
    \caption{Comparison of SSD lifetime and reliability.}
    \label{graph:lifetime}
\end{figure}

To evaluate the lifetime enhancement of \proposal, we measure \mrberx{PEC} for each real tested block, the maximum RBER within the pages in the block, while varying PEC under 1-year retention at \degreec{30}.
We construct five sets of 120 blocks randomly selected from 160 real 3D NAND flash chips and test each set while increasing PEC using one of the five erase schemes.
\fig{\ref{graph:lifetime}} depicts the average \mrber across the tested blocks under different PEC.

\begin{figure*}[t]
    \centering
    \includegraphics[width=\linewidth]{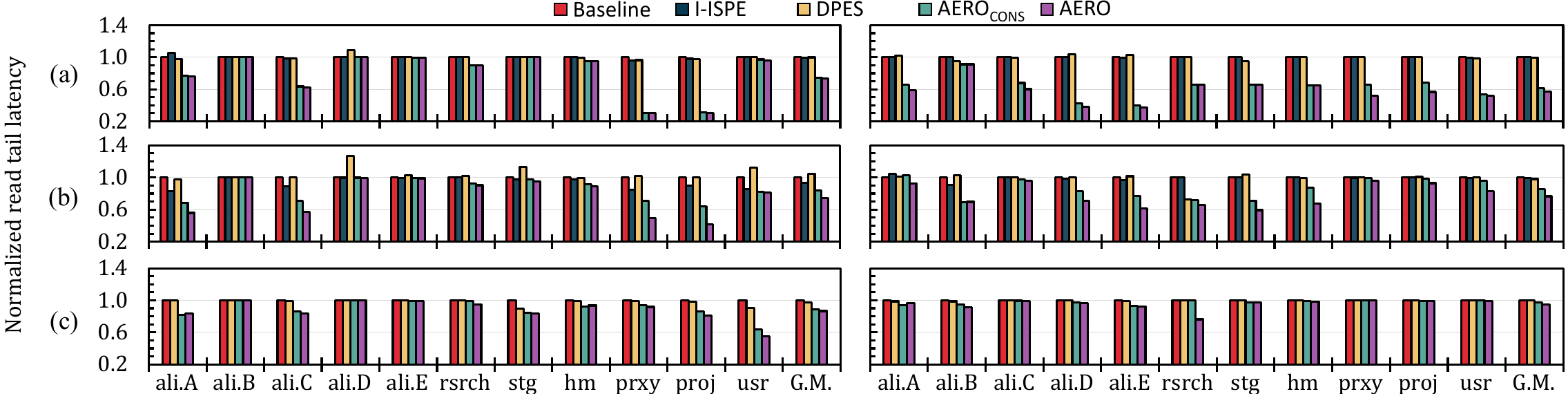}
    \caption{The 99.99th (left) and 99.9999th (right) percentile read latency at $PEC=\langle$(a) 0.5K, (b) 2.5K, (c) 4.5K$\rangle$.}
    \label{graph:lat}
\end{figure*}

We make four key observations. 
First, both \aerooptssd and \aerossd significantly improve SSD lifetime over \basessd.
The average \mrber increases at a much slower rate with PEC in \aerossd and \aerooptssd compared to in \basessd, which clearly shows the high effectiveness of the erase latency reduction in \proposal at lowering erase-induced cell stress.
The slower increase in \mrber, in turn, enables a block to meet the RBER requirement at higher PEC in \aerooptssd (7.6K) and \aerossd (6.9K) compared to \basessd (5.3K), significantly enhancing SSD lifetime by 43\% and 30\%, respectively.

Second, \aerooptssd further improves SSD lifetime considerably over \aerossd (by 10\%) without compromising data reliability.
This highlights the high effectiveness of leveraging the reliability margin to reduce erase latency more aggressively.
The aggressive \tep reduction causes high \mrber even for fresh blocks (i.e., \mrberx{0}$=46$ in \aerooptssd) but greatly slows down \mrber increase (\mrberx{6K}$-$\mrberx{0}$=$9.5), showing its high long-term benefits.

Third, \dpesssd also improves SSD lifetime considerably (by 26\%) compared to \basessd, but its benefits are limited compared to both \aerossd and \aerooptssd.
Like \aerooptssd, \dpesssd exhibits rather high \mrber due to \verase reduction until 3K PEC, which, in turn, enables its \mrber to increase more slowly later.
However, \dpesssd's benefits are limited due to \inum{i}~its limited applicability (until 3K PEC) as well as \inum{ii}~its write-performance overheads.

Fourth, \iispessd \emph{accelerates} RBER increase, which, in turn, rather \emph{decreases} SSD lifetime. 
In fact, \iispessd provides the lowest \mrber among the compared SSDs at $PEC=\langle$0K, 1K$\rangle$ where it is relatively easy to completely erase a block compared to at high PEC.
However, it frequently incurs an erase failure as PEC increases, causing more erase-induced cell stress as explained in \sect{\ref{sec:motivation:limits}}.
Consequently, \iispessd leads to shorter SSD lifetime by 25\% even compared to \basessd, which shows its limited applicability in modern SSDs.

\subsection{Impact on I/O Performance}\label{sec:eval:performance}
\head{Average I/O Performance}
To evaluate the performance impact of \proposal, we first compare \emph{average} I/O performance of the five SSDs in three aspects: \inum{i} read latency, \inum{ii} write latency, and \inum{iii} I/O throughput (IOPS, input/output operations per second).
Table~\ref{tab:avg_perf} summarizes the three average performance values in \iispessd, \dpesssd, \aerossd, and \aerooptssd that are normalized to \basessd, on average across all the workloads.
We observe that all the evaluated SSDs \emph{except for} \dpesssd show almost the same average performance for all the workloads and $PEC=\langle$0.5K, 2.5K, 4.5K$\rangle$.
This is because modern SSDs perform erase operations much less frequently compared to reads and writes as explained in \sect{\ref{ssec:motiv_impact}}.
Unlike the other SSDs, \dpesssd significantly increases average write latency and IOPS at $PEC=\langle$0.5K, 2.5K$\rangle$, i.e., when the DPES scheme is applicable.
Note that we do \emph{not} evaluate \iispessd at 4.5K PEC, as it \emph{cannot} meet the RBER requirement before PEC reaches 4.5K as shown in \fig{\ref{graph:lifetime}}.

\begingroup
\def\arraystretch{0.9}
\begin{table}[h]
\renewcommand\theadfont{\bfseries}
\caption{Comparison of average I/O performance.}
\label{tab:avg_perf}
\centering
\resizebox{\columnwidth}{!}{
\begin{tabular}{cccc}
\toprule
\multirowcell{3.5}{\textbf{Erase Scheme}} & \multicolumn{3}{c}{\textbf{Geomean of Norm. Avg. Perf. at $\textit{\textbf{PEC}} = \langle$0.5K, 2.5K, 4.5K$\rangle$}} \\\cmidrule{2-4}
& \textbf{Norm. Avg.} & \textbf{Norm. Avg.} & \textbf{Norm. Avg.} \\
& \textbf{Read Latency [\%]} & \textbf{Write Latency [\%]} & \textbf{IOPS [\%]}\\
\midrule
\textsf{I-ISPE} & $\langle$100.0, 99.8, N/A$\rangle$ & $\langle$100.0, 100.0, N/A$\rangle$  & $\langle$100.0, 100.1, N/A$\rangle$ \\
\midrule
\textsf{DPES} & $\langle$100.4, 101.3, 99.9$\rangle$ & $\langle$110.8, 135.6, 100.0$\rangle$ & $\langle$95.7, 87.8, 100.0$\rangle$ \\
\midrule
\aerossd & $\langle$99.9, 99.7, 99.7$\rangle$ & $\langle$99.8, 99.9, 99.8$\rangle$  & $\langle$100.2, 100.3, 100.3$\rangle$ \\
\midrule
\aerooptssd & $\langle$99.9, 99.6, 99.7$\rangle$ & $\langle$99.8, 99.8, 99.9$\rangle$ & $\langle$100.2, 100.4, 100.3$\rangle$ \\
\bottomrule
\end{tabular}}
\end{table}
\endgroup

\head{Read Tail Latency}
We evaluate the impact of \proposal on SSD read tail latency that is critical in modern enterprise or datacenter server systems~\cite{decandia-sosp-2007, gunawi-2016-socc}.
\fig{\ref{graph:lat}} represents the 99.99th and 99.9999th percentile read latencies (\four and \six, respectively) in the five simulated SSDs at $PEC=\langle$0.5K, 2.5K, 4.5K$\rangle$ (all values are normalized to \basessd).

We make six observations from \fig{\ref{graph:lat}}.
First, \aerooptssd (\aerossd) significantly reduces \four and \six compared to \basessd by 22\% (18\%) and 26\% (20\%), respectively, on average across all the evaluated workloads and PEC.
Second, \proposal achieves higher performance benefits at lower PEC while still providing considerable performance improvements at high PEC. \aerooptssd (\aerossd) outperforms \basessd by $\langle$35\%, 24\%, 9\%$\rangle$ ($\langle$32\%, 15\%, 7\%$\rangle$) when $PEC=\langle$0.5K, 2.5K, 4.5K$\rangle$,
reducing \four and \six compared to \basessd by
$\langle$26\%, 25\%, 13\%$\rangle$ ($\langle$26\%, 16\%, 11\%$\rangle$) and 
$\langle$43\%, 23\%, 5\%$\rangle$ ($\langle$39\%, 14\%, 2\%$\rangle$)
when $PEC=\langle$0.5K, 2.5K, 4.5K$\rangle$.
This is because \proposal only reduces \tep in \epi{N_\text{ISPE}}, which has a higher impact when \nloop is low.
The high benefits at 0.5K PEC (\fig{\ref{graph:lat}a}) also clearly show the high effectiveness of shallow erasure, given that $N_\text{ISPE}=1$ for most blocks.
Third, \proposal improves I/O performance also when the workload is read-dominant (e.g., ali.E and usr).
This highlights the importance of optimizing the latency of erase operations that dictate read tail latency.
Fourth, at 2.5K PEC, \aerooptssd considerably reduces \four and \six over \aerossd by 
11\% on average (up to 34\% and 22\%, respectively), which shows the effectiveness of leveraging the ECC-capability margin for further \tep reduction.
Fifth, \aerooptssd also reduces \four and \six over \iispessd by 
$\langle$26\%, 20\%$\rangle$ 
and $\langle$43\%, 23\%$\rangle$ 
at $PEC=\langle$0.5K, 2.5K$\rangle$, respectively, on average across all the evaluated workloads. 
At 0.5K PEC, both \aerooptssd and \iispessd can completely erase almost every block via a single loop for which only \proposal can reduce \tbers using shallow erasure. 
Even though \iispessd can also decrease \nloop (and thus \tbers) to less than 2 (7~ms) at 2.5K PEC by skipping the first erase loops, \aerooptssd achieves higher benefits due to aggressive \tep reduction and slower \nloop increase (92\% of the blocks can be erased within two erase loops at 2.5K PEC).
Sixth, \aerooptssd reduces \four and \six over \dpesssd by 22\% and 25\%, respectively, on average across all the evaluated workloads and PEC.
In particular, at 2.5K PEC, \four of \dpesssd rather often \emph{increases} by 5\% compared to \basessd due to the increase in \tprog, whereas \aerooptssd causes no performance degradation but provides 25\% benefits on average across all worloads.

\begin{figure}[h]
    \centering
    \includegraphics[width=\linewidth]{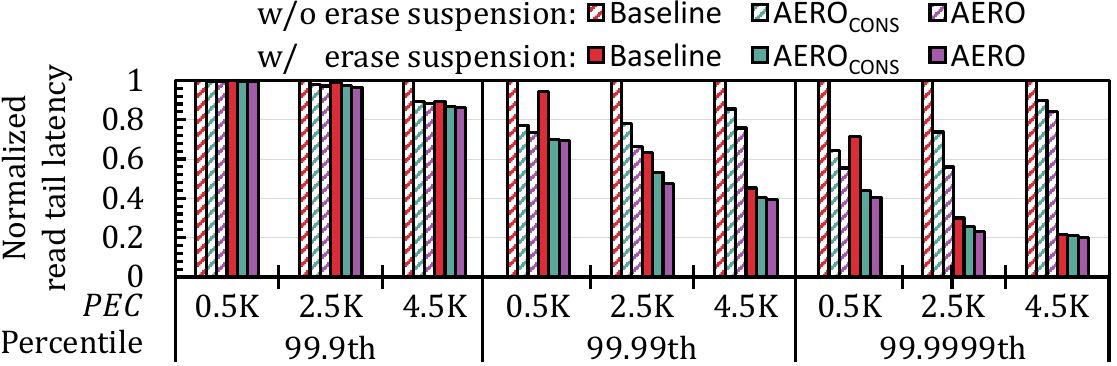}
    \caption{Impact of erase suspension on read tail latency.}
    \label{graph:erase_suspension_percentile}
\end{figure}

\head{Impact of Erase Suspension}
To better understand \proposal's performance benefits, we evaluate the impact of erase suspension on read tail latency in \aerossd and \aerooptssd.
\fig{\ref{graph:erase_suspension_percentile}} shows \three, \four and \six in \aerossd and \aerooptssd at $PEC=\langle$0.5K, 2.5K, 4.5K$\rangle$, averaged across all workloads, when we disable the erase suspension scheme~\cite{kim-fast-2019} (we also plot the results with erase suspension for comparison).
All values are normalized to \basessd without erase suspension.

We make two key observations.
First, \proposal significantly improves I/O performance also when erase suspension is disabled.
\aerooptssd provides high benefits over \basessd, e.g., $\langle$45\%, 44\%, 16\%$\rangle$ reductions in \six at $PEC=\langle$0.5K, 2.5K, 4.5K$\rangle$.
Second, the performance benefits of \proposal become even higher when disabling erase suspension.
For example, when erase suspension is enabled, \aerooptssd achieves $\langle$43\%, 23\%, 5\%$\rangle$ reduction in \six over \basessd at $PEC=\langle$0.5K, 2.5K, 4.5K$\rangle$, which is $\langle$2\%, 21\%, 11\%$\rangle$ lower compared to when erase suspension is disabled.
This is because, without erase suspension, a page read must wait for the completion of the \emph{entire} ongoing erase loop, which significantly increases the impact of erase latency on read tail latency compared to when the ongoing erase loop can be suspended.
Note that, even though the erase suspension scheme often shows a higher reduction in read tail latency to \proposal when they are applied \emph{exclusively}, it does \emph{not} diminish the value of \proposal.
This is because \proposal also improves SSD lifetime significantly and can be easily combined with erase suspension to further reduce read tail latency.

\subsection{Sensitivity Analysis}
\label{sec:eval:sensitivity}
\begin{figure}[h]
    \centering
    \includegraphics[width=\linewidth]{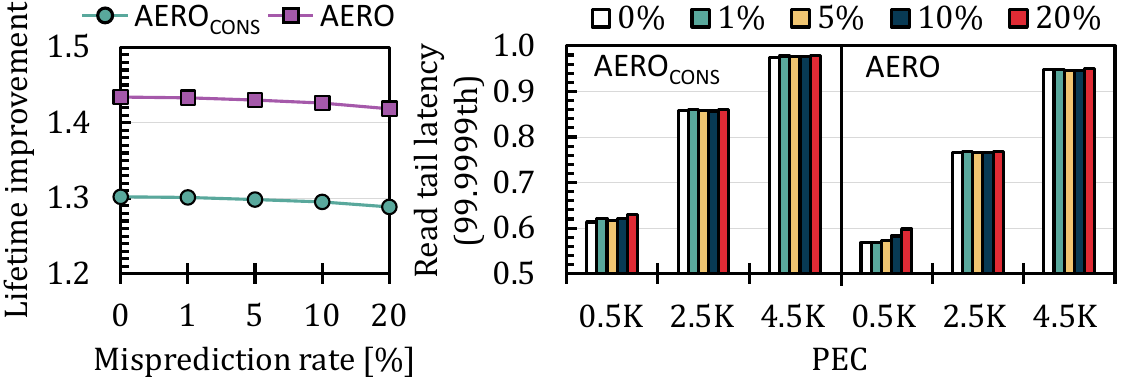}
    \caption{Impact of misprediction rate on \proposal's benefits.}
    \label{graph:sensitivity_misprediction}
\end{figure}

\head{Impact of Misprediction}
Even though we have observed \emph{no} misprediction in our real-device characterization, we analyze the performance and lifetime impact of misprediction in \proposal since it is improbable but not impossible to happen.
To this end, we make two assumptions on \proposal's misprediction behavior based on our real-device characterization results.
First, we consider each erase-latency prediction of \proposal as an independent trial with a constant failure (i.e., misprediction) rate for all blocks and operating conditions.
This is because, although reliability characteristics significantly vary across blocks and operating conditions, \proposal can accurately predict the minimum erase latency for all tested chips, blocks, and operating conditions as demonstrated in \sect{\ref{sec:device_characterization_study}} (i.e., we observe nothing suggesting that certain chips, blocks, or operating conditions are more prone to \proposal's misprediction).
Second, we assume that AERO performs an additional 0.5-ms \ep step for each misprediction.
We believe that 0.5~ms is long enough for \proposal to handle a misprediction because the target block must be largely (though not completely) erased even when a misprediction happens (otherwise, \proposal would not have reduced erase latency).
\fig{\ref{graph:sensitivity_misprediction}} shows how \proposal's misprediction rate affects its benefits in SSD lifetime (left) and read tail latency (right).

We make two key observations.
First, \proposal is highly effective at improving both SSD lifetime and I/O performance even when mispredictions happen.
Even under a high misprediction rate of 20\%, \aerooptssd (\aerossd) provides 42\% (29\%) and 40\% (37\%) improvements over \basessd in SSD lifetime and read tail latency (at 0.5K PEC), respectively.
This is because \aerossd and \aerooptssd are still able to reduce erase latency when a misprediction happens, i.e., the amount of erase-latency reductions (e.g., up to 3.5~ms) is higher than the misprediction penalty (e.g., 0.5~ms), in many cases.
Second, the performance impact of misprediction becomes even smaller as PEC increases.
Compared to when no misprediction occurs, 20\% misprediction rate causes small increases (5.3\% and 2.6\% at 0.5K PEC in \aerooptssd and \aerossd, respectively) in \six, which significantly decreases (to 0.4\% for both) at 4.5K PEC.
This is because the total erase latency severely increases with PEC, thereby making the performance impact of misprediction much smaller.

\begin{figure}[h]
    \centering
    \includegraphics[width=\linewidth]{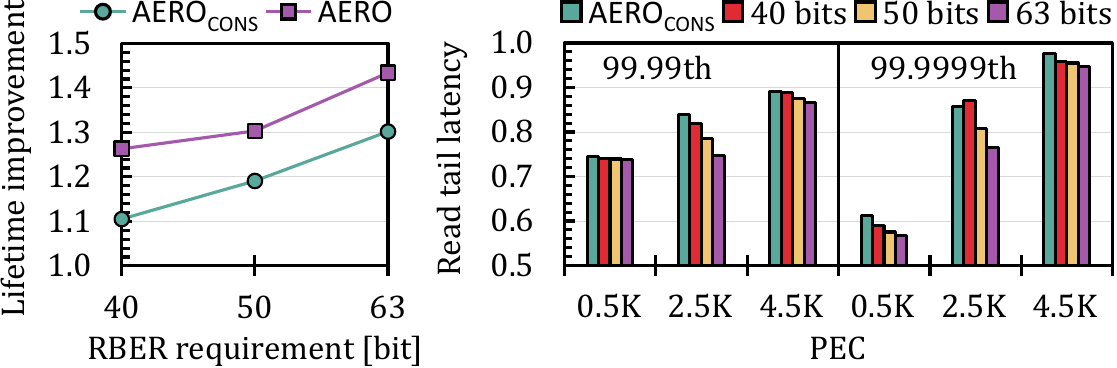}
    \caption{Impact of RBER requirement on \proposal's benefits.}
    \label{graph:sensitivity_ecc}
\end{figure}

\head{Impact of Reliability Margin}
We evaluate how \aerooptssd's benefits change depending on the reliability margin that directly affects the effectiveness of aggressive \tep reduction.
To this end, we evaluate the performance and lifetime benefits of \aerooptssd while reducing the reliability requirement (i.e., the maximum raw bit errors per 1 KiB) to 40 and 50 (from 63), which can happen when using weaker ECC.
\fig{\ref{graph:sensitivity_ecc}} shows SSD lifetime (left) and read tail latency (right) of \aerossd and \aerooptssd under different reliability requirements, normalized to \basessd.
Note that the lifetime of \basessd and \aerossd also degrades as the reliability requirement decreases (because they can tolerate fewer errors).

We observe that \aerooptssd can still improve SSD lifetime when the RBER requirement decreases considerably.
Although the chance for aggressive \tep reduction significantly decreases when the RBER requirement is 40 bits (only if $N_\text{ISPE} = 2$ and \nfail{1} $< \gamma$ as shown in \fig{\ref{graph:rber_opt}b}), it still allows 14\% lifetime enhancement over \aerossd.
In particular, \aerooptssd achieves the highest benefit at 2.5K PEC. 
This is because \aerooptssd can completely erase most blocks with $N_\text{ISPE} \leq 3$, which allows it to aggressively reduce \tep for many blocks ($N_\text{ISPE} \leq 3$ and \nfail{N_\text{ISPE}-1} $< \delta$  in \fig{\ref{graph:rber_opt}b}).

Based on the key observations in our evaluation, we conclude that \proposal is highly effective at improving both SSD lifetime and I/O performance for many real-world workloads under varying operating conditions.
We believe that \proposal is a promising solution, considering its high lifetime and performance benefits that come with almost negligible overheads.
\section{Related Work}
\label{sec:related_work}

To our knowledge, this work is the first to dynamically adjust erase latency by leveraging varying erase characteristics across blocks, providing significant lifetime and performance benefits for modern SSDs.
We already discussed and compared to the state-of-the-art techniques~\cite{ispe,jeong-tc-2017} closely related to \proposal (\sect{\ref{sec:motivation:limits}} and \sect{\ref{sec:evaluation}}).
We briefly describe other related work that aims to improve the lifetime and performance of SSDs.

\head{Mitigating Negative Impact of Erase Operation}
A large body of prior work has proposed various techniques to mitigate the negative impact of erase operations on SSD lifetime and I/O performance.
Many studies have optimized the algorithms of internal SSD management tasks, e.g., garbage collection~\cite{lee-ispass-2011,cui-date-2018,kang-dac-2018,shahidi-sc-2016,choi-hpdc-2018,guo-ipdps-2017,kang-cm-2017,pan-hotstorage-2019,lee-tcad-2013} and wear leveling~\cite{murugan-msst-2011,li-msst-2019,dh-tcad-2022}, to reduce the number of erase operations invoked for servicing the same amount of user writes.
To prevent an erase operation from delaying latency-sensitive reads for a long time, some studies propose to suspend an ongoing erase operation to service user reads (and resume the erase operation after completing the reads)~\cite{wu-fast-2012,kim-fast-2019}.
Despite the significant lifetime and performance improvements made by the prior research, the existing techniques erase a block using the conventional ISPE scheme, thereby causing over-erasure of blocks frequently.
\proposal introduces only small implementation overheads and thus can be easily integrated into the existing techniques to further improve the lifetime and performance of modern SSDs.

\head{Process Variation}
Many prior studies~\cite{yen-hpca-2022, shim-micro-2019, hong-fast-2022, wang-acm-2017, chen-dac-2017, hung-ssc-2015, kim-pe-2021} leverage varying physical characteristics across flash cells to optimize modern SSDs.
Hong~et~al.~\cite{hong-fast-2022} propose a new erase scheme that applies a low voltage to error-prone WLs selectively (while keeping the same voltage for the other WLs), which makes only a small fraction of weak WLs (temporarily) unusable but eventually extends SSD lifetime.
To fully utilize the potential lifetime of NAND flash blocks, Kim~et~al.~\cite{kim-pe-2021} propose a new block wear index that can reflect significant endurance variation across blocks.
Shim~et~al.~\cite{shim-micro-2019} propose to skip some program-verify steps to improve I/O performance if the target WL has better reliability characteristics compared to other WLs.
Out of many process-variation-aware optimizations, to our knowledge, our work is the first to identify a new optimization opportunity to improve both SSD lifetime and I/O performance by leveraging the significant variation in the minimum erase latency across blocks.
\section{Conclusion}
\label{sec:conclusion}
We propose \proposal, a new block erasure scheme that significantly improves both the lifetime and performance of modern NAND flash-based SSDs by dynamically adjusting erase latency.
We identify new opportunities to optimize erase latency by leveraging varying characteristics across flash blocks and the large reliability margin in modern SSDs.
Throughout extensive characterization of 160 real 3D NAND flash chips, we demonstrate that it is possible to \inum{i}~accurately predict the minimum latency just long enough to completely erase a block based on in-execution information (i.e., fail-bit count) and \inum{ii}~aggressively yet safely reduce erase latency by exploiting the reliability margin.
Our results show that \proposal effectively improves SSD lifetime and read tail latency with low overheads for diverse real-world enterprise and data center workloads under varying operating conditions.
\begin{acks}
We thank our shepherd Jian Huang and anonymous reviewers of ASPLOS 2024 for their valuable feedback and comments.
This work was supported by the MOTIE (Ministry of Trade, Industry and Energy) (1415181081), KSRC (Korea Semiconductor Research Consortium) (20019402), NRF (National Research Foundation of Korea) (RS-2023-00283799), and Samsung Electronics Co., Ltd (IO230411-05858-01).
Jisung Park and Myungsuk Kim are the corresponding authors.
\end{acks}
\appendix
\section{Appendix: Terminology Summary}\label{sec:appendix}
Table~\ref{tab:terminology} summarizes new terminologies defined in this work.
\begingroup
\def\arraystretch{1.2}
\begin{table}[h]
\renewcommand\theadfont{\bfseries}
\caption{Summary of newly defined terminologies.}
\label{tab:terminology}
\centering
\resizebox{\columnwidth}{!}{
\small
\begin{tabular}{cc}
\toprule
\thead{Terminology} & \thead{Definition} \\
\midrule
\nloop & Number of erase loops for complete erasure \\
\vri{i} / \epi{i} & $i$-th Verify-Read / Erase-Pulse step \\
\nfail{i} & Number of fail bits after \epi{i} \\
\fpass & Predefined erase pass threshold\\
\fhigh & Full erase pulse threshold\\
\tep{} / \tvr & Erase-Pulse / Verify-Read latency \\
\tshallow{} / \texttt{tRE} & Shallow / Remainder erase latency \\
\mtbers{} / \mtepi{i} & Minimum \tbers{} / \tep(i) \\
\mrber & Maximum raw bit errors \\
\bottomrule
\end{tabular}}
\end{table}
\endgroup

\balance
\bibliographystyle{unsrt}
\bibliography{refs}

\end{document}